\documentstyle[aps,pre,multicols]{revtex}
\input{epsf}
\draft
\begin{document}
\addtolength{\voffset}{0cm}
\title{Grain Dynamics in a Two-dimensional Granular Flow}
\author{S. H{\o}rl{\"u}ck and P. Dimon}
\address{The Center for Chaos and Turbulence Studies\\
The Niels Bohr Institute, Blegdamsvej 17,
DK-2100 Copenhagen {\O}, Denmark\\\vspace{2mm}
(To appear in Phys. Rev. E.)}
%%\date{\today}
\maketitle
\begin{abstract}
We have used particle tracking methods to study the dynamics of individual 
balls comprising a granular flow in a small-angle two-dimensional funnel.  
We statistically analyze many ball trajectories to examine 
the mechanisms of shock propagation. 
In particular, we study the creation of, and interactions between, shock waves.
We also investigate the role of granular temperature and draw 
parallels to traffic flow dynamics.  
\end{abstract}

\pacs{PACS Numbers: 45.70.Mg}

%%\preprint
\begin{multicols}{2}
\narrowtext

\section{Introduction} \label{intro}

Density waves are known to propagate in a variety of granular flow systems.
For example, flows in funnels or hoppers can have slowly propagating 
density waves that may move up or down depending on sand properties 
and geometry\cite{brown}.
They also occur in long pipes\cite{matsushita} and closed 
hourglasses\cite{verveen} as a result of the
interaction with the interstitial fluid.
In particular, they have been observed in a flow of monodisperse balls 
in both small-angle\cite{VD96,HD99} and wide-angle\cite{pennec} 
two-dimensional funnels.  
In the former, one in fact observes kinematic shock waves 
which propagate upstream.

The kinematic shock waves in a small-angle two-dimensional funnel 
have a complex phenomenology of their own and have been studied 
extensively in a previous work~\cite{HD99} (hereafter referred to as HD).
It was established that the flow can be divided into three different 
types depending on the funnel half-angle $\beta$, each having
a different characteristic shock speed $U$:
\begin{itemize}
\item[]{\bf Pipe flow} ($\beta < 0.1^\circ$): the shocks can be
stationary or very slow ($U < 10$~cm/s) but always stop before 
reaching the inlet to the funnel.
\item[] {\bf Intermittent flow} 
($0.1^\circ < \beta < 0.5^\circ$): strong shocks propagate the full
length of the funnel with $50 < U < 150$~cm/s.
\item[] {\bf Dense flow} ($\beta > 0.5^\circ$): the flow is densely
packed and the shocks are weak but fast ($100 < U < 300 $~cm/s).
For $\beta > 2.0^\circ$, shocks can no longer be observed by eye.
\end{itemize} 
In general, it was found that both the average local shock speed $U(x)$ 
and shock frequency $\nu(x)$ increased with the distance from 
the funnel outlet $x$.  
It also appeared that $U(x) \sim w(x)/D$ where $w(x)$ is the 
local funnel width and $D$ is the funnel outlet width. 
Moreover, it was established that the shocks are mainly created 
at specific positions as a consequence of the monodispersity.
Finally, different types of interactions between shocks 
were also observed.

In the present work we simultaneously track all balls in the field of view 
of a camera to study the flow on smaller length and time scales.
This enables us to observe shock creation and interaction more closely.
We can then compute the granular temperature and also traffic flow curves.
The paper is structured as follows. 
In Sec.~\ref{data_acq} we describe the experimental setup and 
the particle tracking method.
In Sec.~\ref{lagr_flow} we study the properties of individual 
ball trajectories.
In Sec.~\ref{euler_flow} we reduce the ball flow to various
one-dimensional fields.
In Sec.~\ref{shockstart} we discuss how shock waves are created
and in Sec.~\ref{interact} we describe their interactions.
In Sec.~\ref{grantemp} we discuss the role of granular temperature
in the flow.
In Sec.~\ref{xavg} we present the time averaged flow properties.
In Sec.~\ref{traffic} we compare the ball flow with traffic flow.
Finally, we summarize our results in Sec.~\ref{summ}.
%% FIG 1:
\begin{figure}  
\epsfxsize=8.5cm
\centerline{\hbox{ 
\epsffile{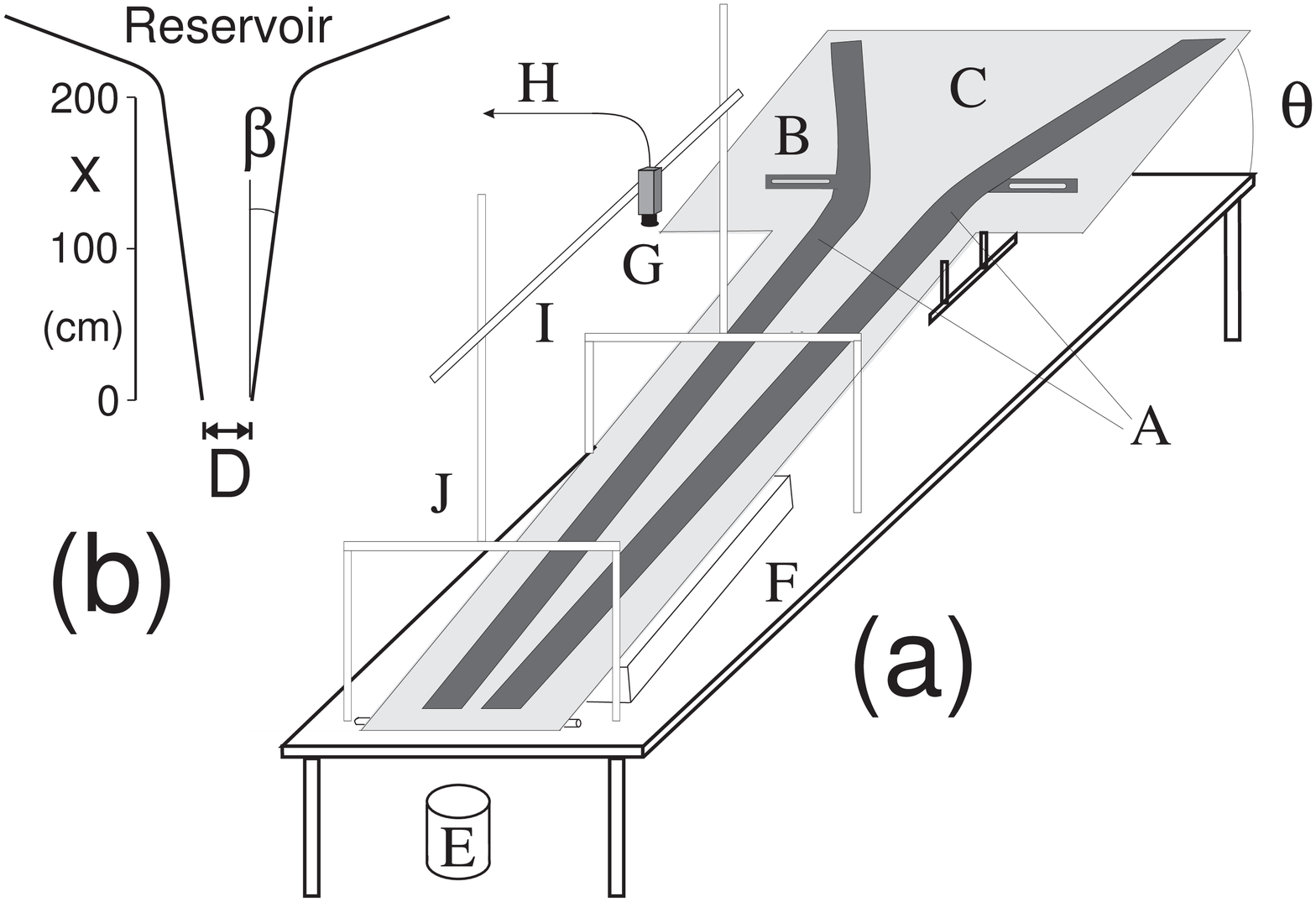}
}}
\caption{ 
(a) Schematic of the experiment. (b) Parameters of the funnel geometry.
}
\label{fig:01} 
\end{figure}

\section{Data acquisition}\label{data_acq}

\subsection{Experimental setup} \label{setup}
%%FIG_01

The experimental setup is described in detail in HD
and will only be briefly reviewed here.
A top view is shown in Fig.~\ref{fig:01}(a). 
The granular matter consists of 50000 3.18~mm diameter brass balls.
They roll in one layer between 3.45~mm high aluminum walls (A) 
on a coated Lexan plane (B). 
The aluminum walls have $200$~cm long straight sections which then 
open smoothly at the top forming a reservoir area (C). 
The straight sections can be moved to vary
the outlet width $D$ and the half-angle $\beta$ of
the funnel (see Fig.~\ref{fig:01}(b)). 
The Lexan plane is tilted an angle $\theta$ so the balls
flow through the funnel into a collection container (E). 
Another plate (not shown) was placed on top of the aluminum 
walls to keep the flow in a single layer. 
Unless stated otherwise, the outlet width and inclination angle were 
kept fixed at $D=10$~mm and $\theta = 4.1^{\circ}$, respectively.
The funnel half-angle was varied in the range $\beta = 0^\circ-3^\circ$.

Elements of the data acquisition system are also shown in 
Fig.~\ref{fig:01}(a).
A light box (F) was placed below the lower $120$~cm 
of the funnel and a video camera (G) was placed above it.
Its analog output signal (H) was
sent to a frame grabber in a PC (not shown). 
The camera was mounted on a stiff bar (I) and could be moved along its length.
The bar was supported by a stabilized stand (J).

\subsection{Video system}

The video camera is a Pulnix TM-6701AN, $8$-bit grey tone, 
non-interlaced, analog CCD camera.
It is capable of filming in four modes of which we used two: 
$640 \times 200$ pixels at $130$ frames/second (fr/s) and
$640 \times 100$ pixels at $221$ fr/s.  
The camera shutter may be adjusted from {\it no shutter} to 
$1/32000$~s.  Shutter speeds of $0.5$-$1$~ms are mostly used 
due to the available light intensity.
The analog signal from the camera is not a standard video signal
but has a pixel rate of 25.5~MHz. 
This signal is transmitted to a Matrix Vision PCimage SGVS frame 
grabber card. 
This PCI-bus PC-card has almost no internal memory and data is 
transmitted by direct memory access to the main memory 
of the host computer.  
Up to 32 MBytes may be grabbed per measurement.
Thus, for $\beta < 0.1^\circ$, up to 2000 consecutive 
frames may be grabbed.  For the largest angles only
$\sim 400$ consecutive frames can be stored. 
Consequently, recorded sequences typically span 2-9~s, 
corresponding to 1-15 shocks depending on $\beta$ and $x$.

The camera is equipped with a 16~mm C-mount lens with a
field of view of $\sim 20^{\circ}$. 
The camera is oriented such that the $640$ pixel direction is
parallel with the center line of the funnel. 
The {\it view length} is used to denote the length of 
the funnel visible in these 640 pixels.  
The lens was set at a height of 106~cm giving a 36.9~cm view length.
The camera was placed so as to cover one of the following ranges: 
$x= -0.2$-36.7~cm, $x= 36.8$-73.7~cm, or $x= 73.8$-110.7~cm.
%%FIG_02
%% FIG 2:
\begin{figure}  
\epsfxsize=8.5cm
\centerline{\hbox{ 
\epsffile{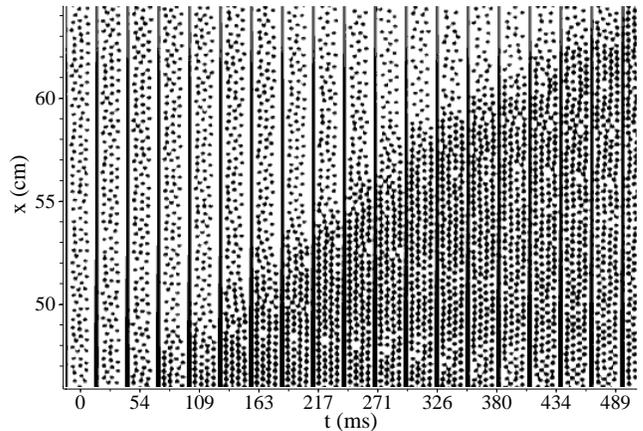}
}}
\caption{
Sequence of video frames showing a propagating shock 
($\beta = 0.15^{\circ}$, $D = 10$~mm, $\theta = 4.1^{\circ}$). 
Only every fourth frame is shown. 
}
\label{fig:02}
\end{figure}

The aluminum walls are painted black and extended with black plates 
to ensure that the only non-black part of the camera's field of view
is the funnel lighted from below. 
The balls therefore appear as black ``dots'' in a white funnel 
as shown in the frame sequence in Fig.~\ref{fig:02}.

\subsection{Ball tracking method} \label{trackingmethod}

At 221~fr/s, a ball speed of 1~ball diameter per frame corresponds 
to a velocity of $\sim 70$~cm/s. 
Dense flows must be slower to be tracked.
In low density flows, tracking is possible at significantly higher 
speeds due to the larger ball separations.
(Computer based particle tracking is the only feasible method 
for data sets containing up to a million ball images.)

Previously, several experiments with fast, low density flows have 
been recorded on high speed film ($\sim 1400$~fr/s), with ball 
positions then obtained by eye~\cite{drake}. 
This method is useful for following relatively small groups of 
particles over relatively short periods of time. 
Warr et al.~\cite{warr} have used expensive digital high speed video 
equipment to follow a few (27-90) fast particles.  
Particle positions are subsequently determined by computer using the 
Hough transformation (an edge detection method~\cite{ballard}) 
for determining ball perimeters.
This method requires many pixels ($\gtrsim 150$) for each ball.
By comparison, we use $\sim 30$~pixels/ball. 

\begin{figure}  
\epsfxsize=8.5cm
\centerline{\hbox{ 
\epsffile{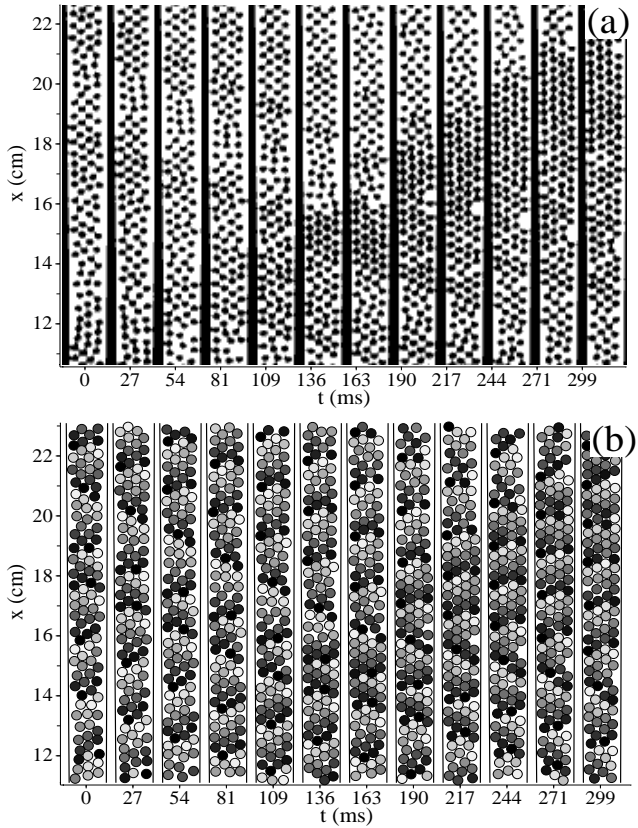}
}}
\caption{
Comparison of (a) video frames and (b) corresponding 
artificial frames using ball tracking 
($\beta=0.3^{\circ}$, $D=10$~mm, and $\theta=4.1^{\circ}$).  
In (b) each ball is identified by a shade of grey.
}
\label{fig:03}
\end{figure}

We now sketch the algorithm used in the present experiment.
(The full details are described in Appendices~\ref{app:ballfind} 
and~\ref{app:pixelif}.)
Each frame is subtracted from the image of the empty funnel. 
Thus, the balls stand out as a white patch on a black background. 
A threshold is then determined based on local grey scale 
distributions and everything above the threshold is then 
defined as part of a ball. 
Each ball center position and height (i.e., grey scale intensity) 
is estimated, and a number of pixels are assigned to
the ball image (typically 20-35). 
These data are used as input to a fitting routine. The pixel values 
(with coordinates $(x,y)$) are fitted to a two-dimensional peak
function and the center of the fitted peak position determines
the ball center ($x^j_{c}(t_k)$,$y^j_{c}(t_k)$) for the $j$th ball
in the $k$th frame.
The final result is illustrated in Fig.~\ref{fig:03}.
%%FIG_03
%% FIG 3:

Each frame has now been converted into a table of ball positions. 
Based on the correlations between the first two frames and their 
densities, an initial guess is made of the one-dimensional 
coarse-grained {\em velocity field} $v(x,t_1)$ of the first frame. 
The $j$th ball in the first frame is now assigned a velocity
$(v^j_x(t_1),v^j_y(t_1)) = (v[x^j_c(t_1),t_1],0)$.
In the interest of clarity, we will henceforth drop the indices
$j$ and $k$.
Based on these velocities, predictions of the possible positions in the 
next frame are determined and compared with the actual ball positions. 
Positions are matched ``one-to-one'' making the best matches first. 
Balls in the old frame that cannot be matched acceptably are defined as lost. 
Matching is first attempted at the position predicted for constant
$(v_x(t),v_y(t))$, then at the old position 
$(v_x(t),v_y(t)) = (0,0)$, and finally at a midway point.  
Unmatched new balls are assigned $(v_x(t),v_y(t) = (v(x,t),0)$ 
at their respective positions. 
(In each frame $v(x,t)$ is re-computed from averages of those balls 
which have known velocities.)
By repeated application of this method, the full trajectory
of each ball can be found. 

The error of $(x_c,y_c)$ is $\lesssim 0.15$~mm, but in a 
few bad cases it may be as much as $\sim 0.5$~mm. 
Less than $0.05\%$ of the balls are normally lost in each frame.
Most balls can be followed from the time they enter the 
view length of the camera until they leave it.  
Bad matches are very rare. 
(They only seem to happen in groups and are easily recognized.)
 %%FIG_04
%% FIG 4:
\begin{figure}  
\epsfxsize=8.5cm
\centerline{\hbox{ 
\epsffile{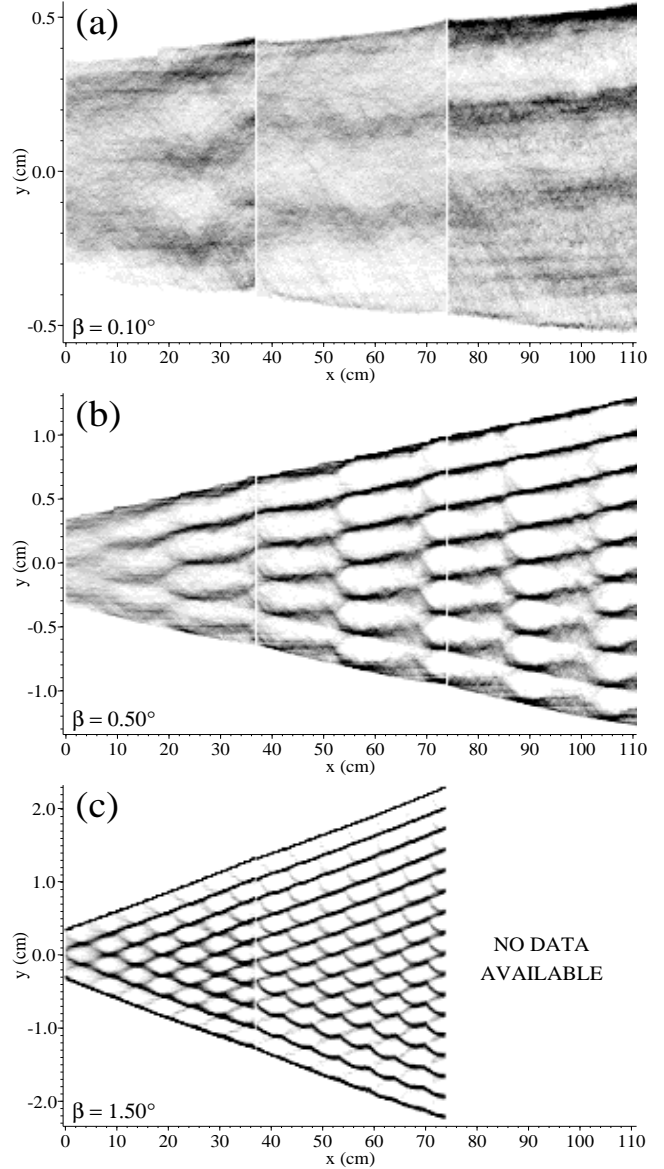}
}}
\caption{
Histograms of measured ball center positions. 
(a) $\beta = 0.1^{\circ}$. One only observes weak structures.
(b) $\beta = 0.5^{\circ}$. There are now $4 - 10$ triangular packing rows.
(c) $\beta = 1.5^{\circ}$. There are $4 - 17$ triangular packing rows.
(The vertical white stripes separate independent data sets.
The weak horizontal stripes in (a) and (b) with a separation of
$\sim 0.6$~mm  are caused by interference with camera pixel rows.
See Appendix B.)
}
\label{fig:04}
\end{figure}
 
The final determination of the velocity ($v_x(t)$,$v_y(t)$) and 
acceleration ($a_x(t)$,$a_y(t)$) of each ball is based on a least squares 
fit of the positions to a parabola using a number of consecutive points. 
This number represents a tradeoff between precision and time resolution. 
For velocities, 5 points are generally used, resulting in a 
time resolution of $\sim 23$~ms. 
For accelerations, 5 or 7 points are used depending on the application.
The latter corresponds to a time resolution of $\sim 32$~ms.

\section{Flow in Lagrangian coordinates}\label{lagr_flow}

\subsection{Single ball trajectories}\label{trajectories}

Since each measurement typically involves 0.3-1.0~million balls, 
it is possible to measure high resolution distributions of
the ball center positions ($x_c(t)$,$y_c(t)$) (averaged over all frames).
Some examples are shown in Fig.~\ref{fig:04}.
For intermittent flow (Fig.~\ref{fig:04}(a)) we observe
weak periodic patterns across the funnel. For dense flows
(Figs.~\ref{fig:04}(b) and (c)), these patterns are much more 
pronounced.  (No such patterns are observed for pipe flow.)

As discussed in HD, the monodispersity of the balls allows 
triangular close packing at packing sites $x=\chi_i$ given by
\begin{equation}  
\label{eq:packsites}
\chi_i = \frac{ 2r + \sqrt{3} r (i-1) - D }{2 \tan \beta }
\end{equation}
where $i$ is an integer and $r$ is the ball radius.
This predicts that such packing sites should recur 
at intervals of $\sqrt{3} r$ and $\sqrt{3} r/2 \tan \beta$ in the 
funnel axis and transverse directions, respectively.
This is what is indeed observed as can be seen in 
Figs.~\ref{fig:04}(b) and~(c) to within $\sim 3\%$.
Immediately downstream of a packing site the lattice must rearrange itself. 
It appears that this occurs over a relatively short distance.
For example, in Fig.~\ref{fig:04}(b) where ${\chi}_{7}=56$~cm and
${\chi}_{8}=72$~cm, the rearrangement occurs at $x=70$~cm.
As one might intuitively suspect, shock waves are more readily  
produced where collisions are most likely.
We believe this is the reason for the nearly exclusive creation of 
shock waves at the packing sites $\chi_i$ as discussed in HD.

\begin{figure}  
\epsfxsize=8.5cm
\centerline{\hbox{ 
\epsffile{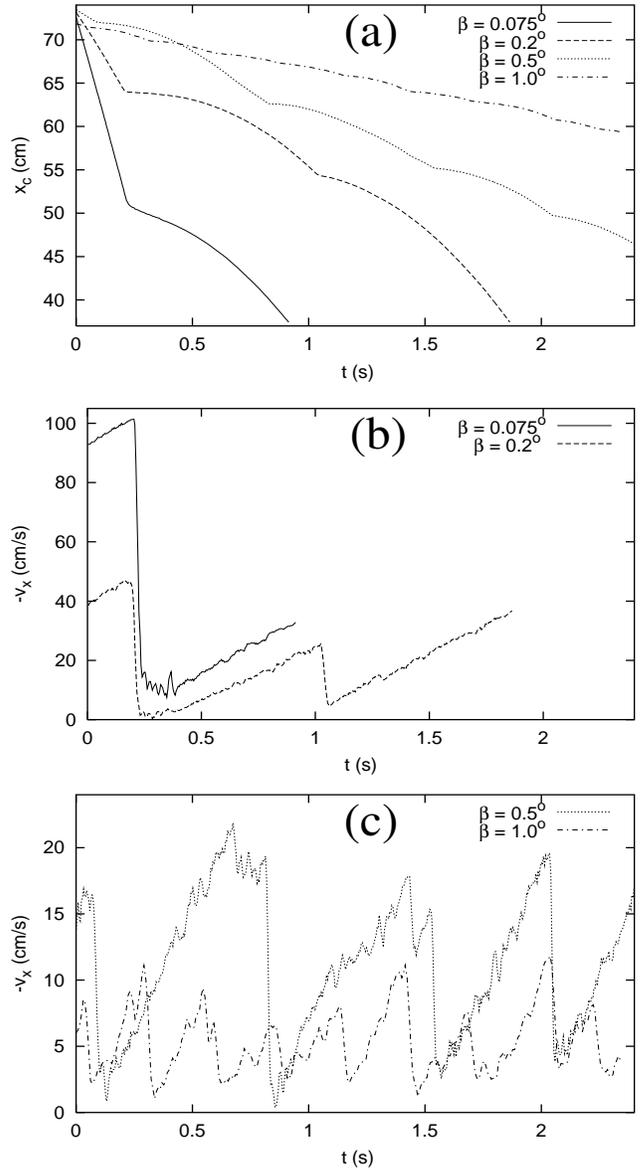}
}}
\caption{
(a) The trajectories of four balls for different funnel angles.
The $x_c(t)$-coordinate is plotted for each ball.
(b) The velocity $v_x(t)$ of the two lower trajectories in (a)
(small funnel angles).
(c) The velocity $v_x(t)$ of the two upper trajectories in (a)
(large funnel angles).
}
\label{fig:05}
\end{figure}

Fig.~\ref{fig:05}(a) shows four representative examples of individual ball 
trajectories obtained using the method described in Sec.~\ref{trackingmethod}. 
Figs.~\ref{fig:05}(b) and (c) show $v_x(t)$ for the same four balls.
Note that since the balls almost always move downstream, 
we always plot $-v_x(t)$ for convenience.
It is apparent from these figures that an individual ball 
experiences long periods of constant (negative) 
acceleration interrupted by abrupt drops in speed to
$|v_x| \sim 1-5$~cm/s for $\beta > 0.2^{\circ}$ and 
$\sim 5-15$~cm/s for $\beta < 0.1^{\circ}$. 
These drops occur when the ball encounters a shock.
This qualitative behavior is even more apparent in the acceleration $a_x(t)$.  
In Fig.~\ref{fig:06} we show the velocity $v_x(t)$ (a) and 
acceleration $a_x(t)$ (b) for another trajectory. 
%%FIG_05
%% FIG 5:

\subsection{Collisions and the coefficient of restitution} 
\label{restitut}

By observing collisions between two balls, we can measure
(with limited resolution) the coefficient of restitution 
$\varepsilon$ of a ball, defined as the ratio of the relative
velocities before and after a collision.
This can only be done for flows with low densities and no 
shock waves in the field of view so that individual collisions 
can be identified. 
In two data sets at $\beta = 0^{\circ}$ that adhere to these
requirements, we have measured a total of 192 collisions and
found that $\varepsilon = 0.78 \pm 0.02$. 
The average relative velocity before these collisions was
$v_{rel} = 7 \pm 3$~cm/s, compared with the actual ball 
speed $|v_x| = 112 \pm 6$~cm/s. 
Assuming that the balls roll without slipping, the tangential 
velocity difference immediately preceding each collision should be
twice the ball speed and therefore much higher than $v_{rel}$.
The measured value of $\varepsilon$ using this method 
agrees with the value $\varepsilon \gtrsim 0.74$ found
by dropping a ball on a hardened steel block. 

From measurements of fast rolling collisions 
($v_x \sim 40$-80~cm/s) with a steel barrier, 
we estimate $\varepsilon \sim 0.5$-0.6.
The additional energy loss here may come from the rolling ball sliding 
on the contact surfaces at the moment of collision. 
This may be related to the situation where a fast ball encounters 
the slowly moving packed region of a shock wave.
 
\subsection{Shock profiles}\label{profile}

\begin{figure}  
\epsfxsize=8.5cm
\centerline{\hbox{ 
\epsffile{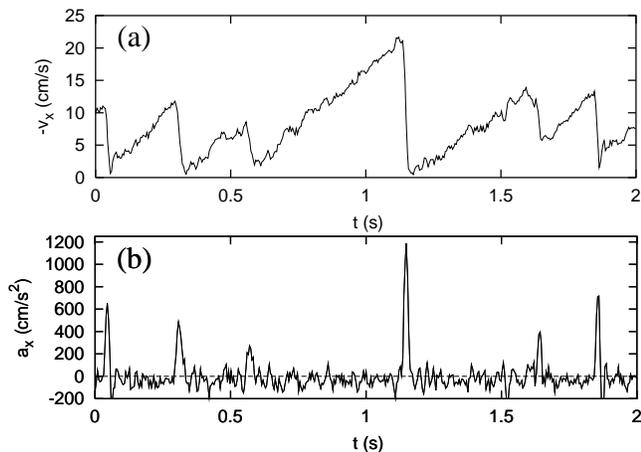}
}}
\caption{
Movement of a single ball at $\beta= 0.8^{\circ}$ between $x=73$~cm and
$x=37$~cm showing the
(a) velocity $v_x(t)$ and (b) acceleration $a_x(t)$.  
The peaks in (b) indicate that the ball has encountered and is
nearly stopped by a shock.  The dashed line marks zero acceleration. 
}
\label{fig:06}
\end{figure}

We showed in Figs.~\ref{fig:05}(b,c) and \ref{fig:06}(a) the 
velocity $v_x(t)$ of single ball trajectories. 
Their apparent sawtooth behavior is, in fact, a general feature 
observed in many trajectories. 
From this simple structure we can obtain some 
interesting information concerning the dissipation of the flow 
between shock waves.  
%%FIG_06
%% FIG 6:

First, we determine the inter-shock acceleration $a_{is}$, 
i.e., the acceleration between shock waves.
Using the acceleration $a_x(t)$ shown in Fig.~\ref{fig:06}(b)
and employing a threshold criterion, we can identify shock waves
and the time $t_s$ when a given ball encounters a shock wave. 
Linear fits yield typical values of $a_{is} = -40 \pm 10$~cm/s$^2$.
This value should be compared with the expected effective 
gravitational acceleration (for spheres rolling without slipping)
$g_r = \frac{5}{7} g \sin{\theta} = -50$~cm/s$^2$ for $\theta=4.1^\circ$.
It is not surprising that $|a_{is}| < |g_r|$ since energy is 
probably lost in collisions and friction processes. 
%%FIG_07
%% FIG 7:
\begin{figure}  
\epsfxsize=8.5cm
\centerline{\hbox{ 
\epsffile{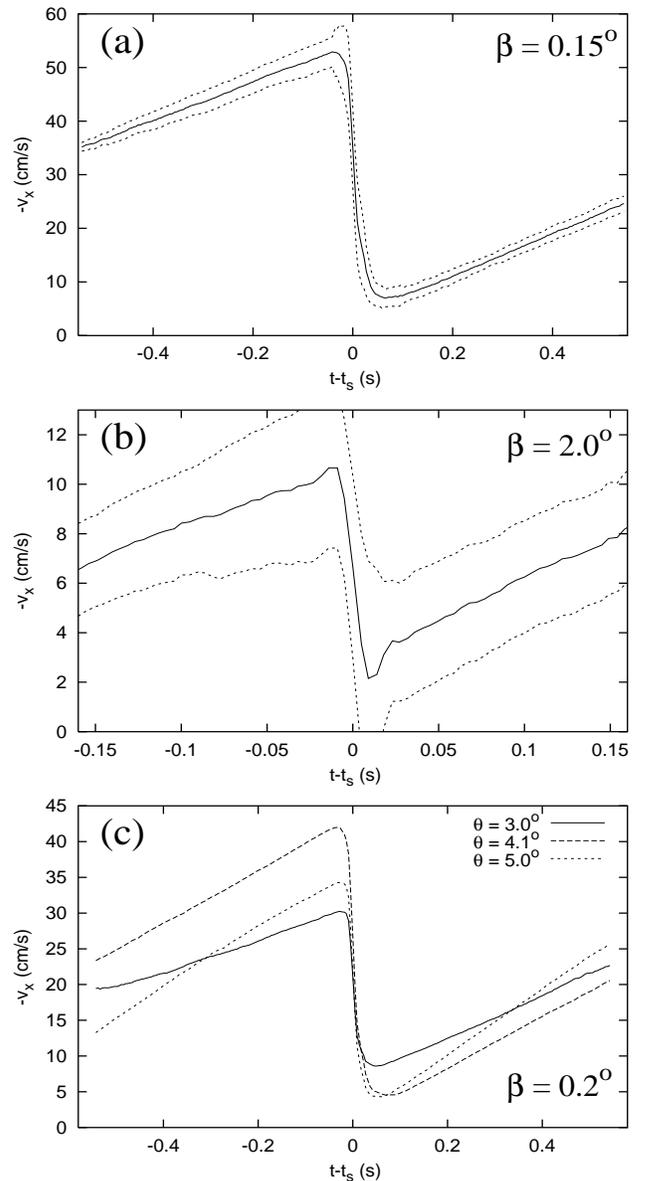}
}}
\caption{
Average velocity profiles of shocks in Lagrangian coordinates.
Specifically, we plot the average of $-v_x(t-t_s)$ of individual balls, 
where $t_s$ is a pronounced local maximum in ball acceleration.
(a) $\beta=0.15^{\circ}$.  Average of one shock.
(b) $\beta=2.0^{\circ}$.  Average of $\sim 10$ shocks.
(c) single shock profile for $\beta=0.2^{\circ}$ and 
different inclination angles $\theta$.
The dashed lines in (a) and (b) indicate the standard deviation 
of the measurement.
}
\label{fig:07}
\end{figure}

Better statistics can be obtained by averaging over many 
trajectories of balls encountering the {\em same} shock wave.
By using a film sequence such as that in Fig.~\ref{fig:02},
we choose a time interval such that only one shock is visible
in the view length.
Single ball trajectories with values of $t_{s}$ within
that interval are then selected (typically a few hundred).
If we now average the trajectories for all these balls
by setting $t_s=0$ for each ball, we obtain average profiles
such as those shown in Figs.~\ref{fig:07}(a) and (b) for
intermittent and dense flow, respectively. 
These preserve the sawtooth structure discussed earlier, reinforcing our 
picture of the behavior of single balls passing through shock waves. 
In particular, the inter-shock acceleration calculated for points
outside the shock region for $\theta = 4.1^\circ$ is 
$a_{is} = -38 \pm 3$~cm/s$^2$.  
It was found to be independent of both velocity 
(i.e., the same just preceding or following a shock wave) 
and $\beta$ when $\beta > 0.1^\circ$ (i.e., for intermittent and dense
flows; there are exceptions in pipe flow when $|v_x| \gtrsim 80$~cm/s).
It was also checked that $a_{is}$ was the same in different regions 
of the funnel, and hence independent of $x$ as it should be.
%%FIG_08
%% FIG 8:
\begin{figure}  
\epsfxsize=8.5cm
\centerline{\hbox{ 
\epsffile{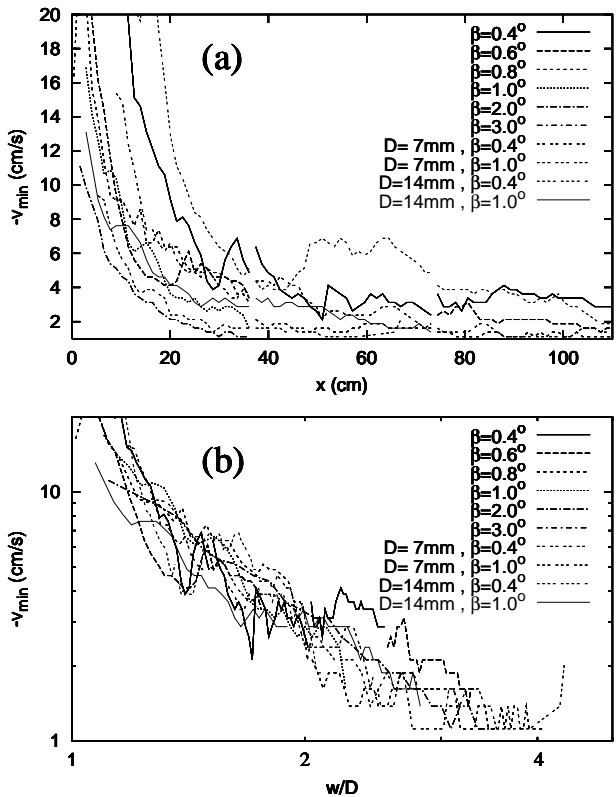}
}}
\caption{
(a) The median of the distribution of $v_{min}$ versus $x$.
(b) The same data replotted against the rescaled variable $w(x)/D$.
}
\label{fig:08}
\end{figure}

We have also checked the dependence of $a_{is}$ on $\theta$.
The velocity profiles for three different values 
of $\theta$ are shown in Fig.~\ref{fig:07}(c). 
Ignoring for now differences in the velocity jumps across
the shock, one sees that $a_{is}$ increases with $\theta$.
For $\theta=3.0^{\circ}$, we find $a_{is} =-27 \pm~3$~cm/s$^2$
and for $\theta=5.0^{\circ}$, $a_{is} =-45 \pm~3$~cm/s$^2$.
(These values were also independent of velocity and $\beta$.)
It appears therefore that $a_{is}/g_r \sim 0.75$ in all
three cases, implying that $a_{is}$ can be regarded as an effective
gravitational acceleration.

Experiments involving only a single ball for $\theta=4.1^{\circ}$ 
yielded a measured value $g_{r} = -45 \pm 5$~cm/s$^2$.
The reason for the difference between this and the theoretical value 
$g_{r} = -50$~cm/s$^2$ is partially due to a local deviation 
of $\theta$ of $\sim 0.2^\circ-0.3^\circ$ from a small 
bending of the plane, and possibly also to non-ideal rolling. 
Consequently, we believe that the difference in $a_{is}$ and the measured 
value of $g_{r}$ is significant, i.e., due to the interaction
between balls in intermittent and dense flows.

Just before and after a shock, a ball reaches its maximum and minimum
velocity $v_{max}$ and $v_{min}$, respectively
(see, for example, Fig.~\ref{fig:06}).
In particular, knowledge of $v_{min}$ is necessary to understand the
creation of shock waves discussed in Sec.~\ref{interact}.
In the same manner as discussed earlier, we find that the average value of 
$v_{min}$ for all balls passing through a given shock wave is
$|v_{min}| \sim 10-20$~cm/s, $4-10$~cm/s, and $1-5$~cm/s, 
in pipe, intermittent, and dense flow, respectively
(see for example Figs.~\ref{fig:05},~\ref{fig:06}, and~\ref{fig:07}).

We also checked the dependence of $v_{min}$ on $\beta$ and $x$.
(Due to the presence of shock waves that are created within the
view length and a low number of shock waves, $v_{min}(x)$ is 
obtained as the median of the distribution of $v_{min}$, not the mean.)
This is shown in Fig.~\ref{fig:08}(a) for 
$ 0.4^{\circ} < \beta < 3.0^{\circ}$ and $D=7$, 10, and 14~mm.
(There were too few shock waves for smaller $\beta$ to obtain
meaningful statistics.) 
In HD it was found that the local shock speed $U(x)$ 
depended linearly on the local funnel width $w(x)$.
If we plot $v_{min}$ against $w(x)/D$, 
we obtain the plot shown in Fig.~\ref{fig:08}(b).
The apparent collapse of the data indicates that there may
indeed be some simple scaling properties inherent in the system.
It will be shown in a future work how the behavior of
$a_{is}$ and $v_{min}$ can be used to model shock wave 
statistics\cite{modelpaper}.

\section{Flow in Eulerian coordinates} \label{euler_flow}

\subsection{Granular fields} \label{define_fields}

From the full set of ball trajectories in the flow, we can construct
continuous one-dimensional Eulerian fields, such as the 
density $\rho(x,t)$ and velocity $v(x,t)$.
(The justification for considering a one-dimensional flow is
given in Appendix~\ref{app:yprof}.) 
Of course, to obtain meaningful results, it is necessary to coarse-grain.
In order to obtain a spatial resolution better than one ball diameter,
this was done by drawing a primitive unit cell 
(for close-packed circles) around the ball center positions 
$(x_c(t),y_c(t))$, with two sides parallel to $x$
(i.e., the center line of the funnel).
The funnel was then binned into strips of width $\sim 0.1$~mm
and the occupied area was calculated for each strip.  
These strips can then be summed to achieve any desired resolution, 
although typically we used a resolution of $\sim 1$~mm.
We then divided by the area of the strip to obtain the relative
density $\tilde{\rho}(x,t)$.
Thus, for close packed balls, $\tilde{\rho}(x,t) = 1$.
(The relative density here differs from that in HD which was
measured using relative light intensities.)
The velocity field $v(x,t)$ is then computed as the weighted average 
\begin{equation}\label{eq:rhoavg}
v(x,t) = \sum_j \tilde{\rho}^j v_x^j(t)/\sum \tilde{\rho}^j 
\end{equation}

\end{multicols}
%%\begin{multicols}{1}
\widetext
%% FIG 9 (Double column):
\begin{figure}  
\epsfxsize=17.5cm
\centerline{\hbox{ 
\epsffile{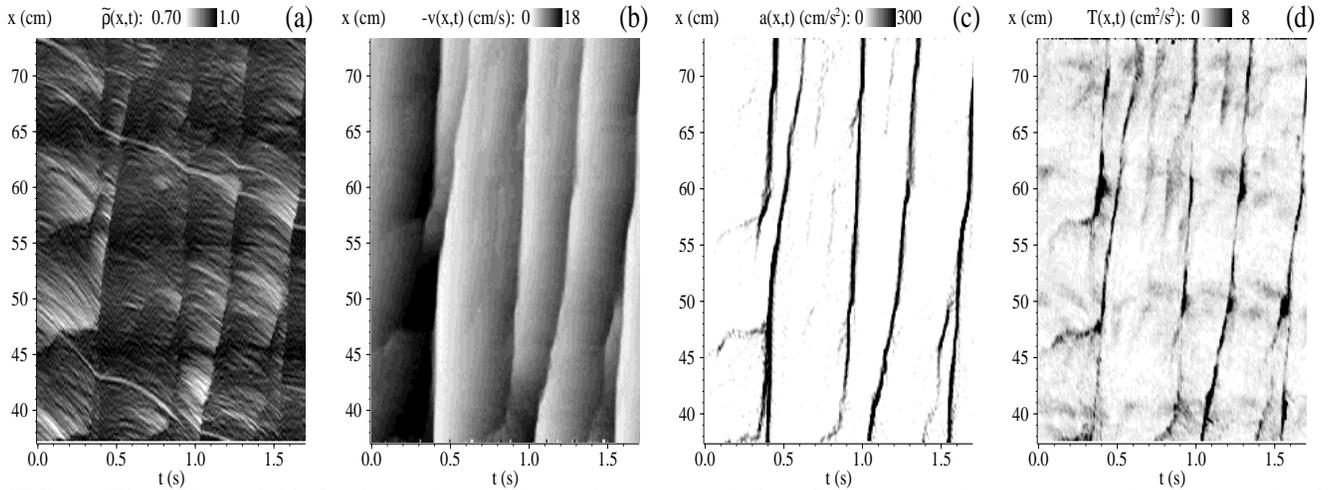}
}}
\caption{
The different fields for $\beta=0.8^{\circ}$.
(a) relative density $\tilde{\rho}(x,t)$,
(b) velocity $v(x,t)$, 
(c) positive acceleration $a(x,t)$, 
(d) granular temperature $T(x,t)$.
}
\label{fig:09}
\end{figure}
\begin{multicols}{2}
\narrowtext

where the sum is over all balls whose unit cells contribute
to a strip and $\tilde{\rho}^j$ is the corresponding relative density.
The acceleration field $a(x,t)$ is obtained in a similar manner. 

The granular temperature per ball is defined as
$T(x,t) = \langle (v_y -\bar{v}_y)^2 \rangle 
+ \langle (v_x - \bar{v}_x)^2 \rangle$ 
where $\bar{v}_x = v(x,t)$ and  $\bar{v}_y = 0$, and where the brackets
indicate an average as in Eq.~\ref{eq:rhoavg}.
It can be written as the two components  
$T(x,t) = {T_x}(x,t) + {T_y}(x,t)$, 
where ${T_x}(x,t) =  \langle [v_x - v(x,t)]^2 \rangle$
and ${T_y}(x,t) =  \langle v_y^2 \rangle$, in order to
see if the granular temperature is isotropic.

\subsection{Discussion of the fields} \label{compare_fields}

An example of the relative density is shown in 
Fig.~\ref{fig:09}(a) for dense flow ($\beta=0.8^{\circ}$).
It reveals five faint shocks. 
However, in dense flows, density fluctuations on the scale of 
a ball diameter tend to be comparable with or even dominate the
fluctuations associated with the passing of shock waves.
(For example, the whiter tracks moving downstream are
defects, which are often vacancies.
In dense flows they can survive the passing of shock waves
since shocks do not seriously rearrange the local packing.) 

The corresponding velocity and acceleration fields are
shown in Figs.~\ref{fig:09}(b) and (c), respectively.
In the velocity field, the same shock waves are much
more clearly visible since the greatest contrast occurs when 
fast balls enter a shock wave and lose most of their speed.
(Note that again, since $v(x,t)$ is always negative, we plot 
$-v(x,t)$ here and in all subsequent plots.)

By adjusting the resolution in the positive acceleration field 
$a(x,t)$ shown in Fig.~\ref{fig:09}(c), we can improve the 
contrast in $v(x,t)$ in order to reveal weaker details of the flow. 
For example, one can now see the creation of (temporarily) 
stationary shocks at locations corresponding approximately
to the packing sites $\chi_8 = 45$~cm and $\chi_9 = 55$~cm
(see Eq.~\ref{eq:packsites}).

The granular temperature $T(x,t)$ shown in Fig.~\ref{fig:09}(d)  
clearly peaks in the shock regions as it did for $a(x,t)$, 
although it is not as evenly distributed along the shocks.
The temperature appears to cool down over relatively 
short time and length scales.
There is a background level of $\sim 2$~cm$^2/$s$^2$ 
with a weak packing site periodicity of $\sim 10$~cm.
The granular temperature will be discussed more thoroughly 
in Sec.~\ref{grantemp}.

\subsection{Flow dynamics for fixed $x$} \label{const_x}

\begin{figure}  
\epsfxsize=17.5cm
\centerline{\hbox{ 
\epsffile{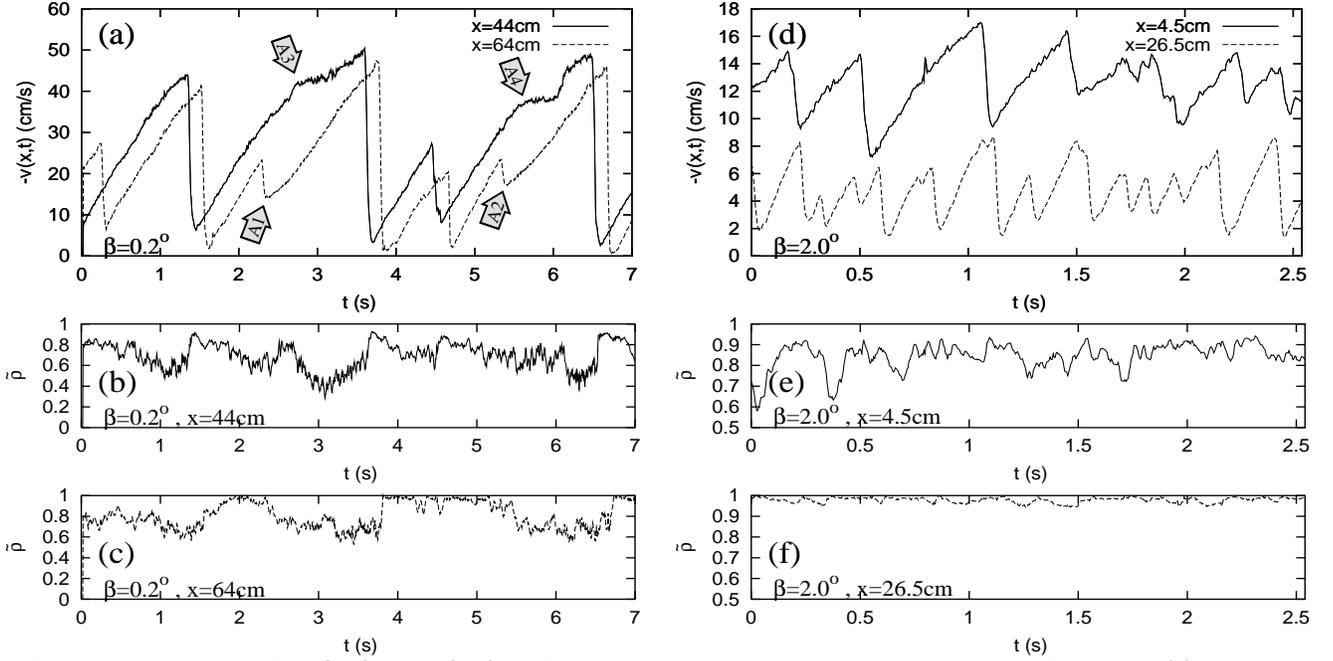}
}}
\caption{
Time series of $-v(x,t)$ and $\tilde{\rho}(x,t)$ at fixed $x$ 
showing propagating shocks for two different $\beta$.
(a) velocity for $\beta=0.2^{\circ}$ at $x=44$~cm (between $\chi_4$ and
$\chi_5$) and $x=\chi_5=64$~cm.
A1 and A2 indicate newly created shocks, and A3 and A4 show, 
respectively, the effect of these shocks 20~cm downstream.
(b) density corresponding to (a) at $x=44$~cm.
(c) density corresponding to (a) at $x=64$~cm.
(d) velocity for $\beta=2.0^{\circ}$ at $x=4.5$~cm (between packing sites) 
and $x=26.5$~cm (at a packing site). 
(e) density corresponding to (d) at $x=4.5$~cm. 
(f) density corresponding to (d) at $x=26.5$~cm.
}
\label{fig:10}
\end{figure}

The grey scale plots of the granular fields in Fig.~\ref{fig:09}
do not convey the magnitudes of the fields very well. 
Therefore we show $v(x,t)$ and $\tilde{\rho}(x,t)$
for four fixed $x$ in Fig.~\ref{fig:10} and study 
how these quantities are affected by propagating shock waves.

In Fig.~\ref{fig:10}(a), the velocity $v(x,t)$ is shown at 
$x=44$~cm (midway between ${\chi}_4 =24$~cm and 
${\chi}_5 =64$~cm), and at $x=\chi_5=64$~cm for an 
intermittent flow ($\beta=0.2^\circ$). 
(These positions were chosen for reasons which will become clear later.)
One can clearly see propagating shock waves and smoothly increasing 
flow speeds between the shocks in the range 3-45~cm/s. 
From the displacements of the shock fronts, we estimate
that $U \sim 120$~cm/s. 
At $x=64$~cm, one can see two shocks labelled A1 and A2.
From fields such as those in Fig.~\ref{fig:09},
we know that these are newly created shocks.
At $x=44$~cm, disturbances are observed at A3 and A4 as the velocity
increases between shocks.
These are caused by balls that have passed the newly created 
shocks at A1 and A2. 
The corresponding relative densities are shown in Figs.~\ref{fig:10}(b,c).
They reveal that the shock fronts are relatively sharp,
but otherwise details of the flow are generally not as clear as in $v(x,t)$.  
\end{multicols}
%%\begin{multicols}{1}
\widetext
%%FIG_10
%% FIG 11 (Double column):
\begin{figure}  
\epsfxsize=17.5cm
\centerline{\hbox{ 
\epsffile{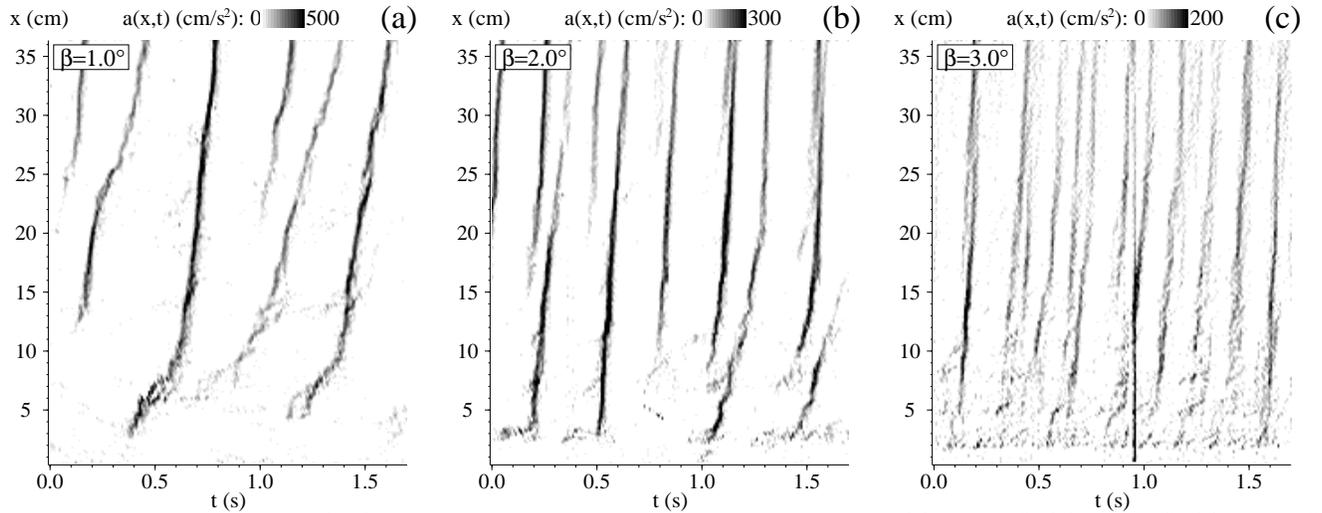}
}}
\caption{
Acceleration field $a(x,t)$ showing weak shock waves for 
large values of $\beta$: (a) $\beta=1.0^{\circ}$,
(b) $\beta=2.0^{\circ}$, (c) $\beta=3.0^{\circ}$.
The grey scale for the magnitude is shown at the top of each plot.
}
\label{fig:11}
\end{figure}
\begin{multicols}{2}
\narrowtext
In Fig.~\ref{fig:10}(d), we show $v(x,t)$ at $x=4.5$~cm
(between ${\chi}_4 =2.4$~cm and ${\chi}_5 =6.5$~cm),
and $x=\chi_{10}=26.5$~cm for a dense flow ($\beta=2.0^{\circ}$).
Many shocks with speeds $U \sim 350$~cm/s are now visible.
At $x=4.5$~cm they are all newly created. 
These shocks can still be observed at $x=26.5$~cm 
along with many other shocks created upstream of them.
The corresponding relative densities are shown in 
Figs.~\ref{fig:10}(e) and (f), respectively. 
These are of limited value when $\beta > 1.5^\circ$
Near the outlet (e) there are still fluctuations related to shock
waves, but upstream (f) the density is constant and nearly
unity despite the presence of many shock waves.

\subsection{Shock behavior for $\beta > 1.0^\circ$} \label{highbeta}

As $\beta$ is increased, the shock waves are faster, weaker,
and more frequent as discussed in HD. 
For $\beta>1.5^{\circ}$, one can no longer study shock
waves from directly measured density fluctuations as in HD and
in this situation the ball tracking method is particularly useful.

In Figs.~\ref{fig:11}(a-c) we show  $a(x,t)$ near the outlet
for three relatively large values of $\beta$. 
One can see that as $\beta$ increases, the frequency and speed 
of the shocks increase while their strength decreases.
In particular, in Fig.~\ref{fig:11}(c) it also appears
that the shock waves die off noticeably with increasing $x$.
%%FIG_12
%% FIG 12:
\begin{figure}  
\epsfxsize=8.5cm
\centerline{\hbox{ 
\epsffile{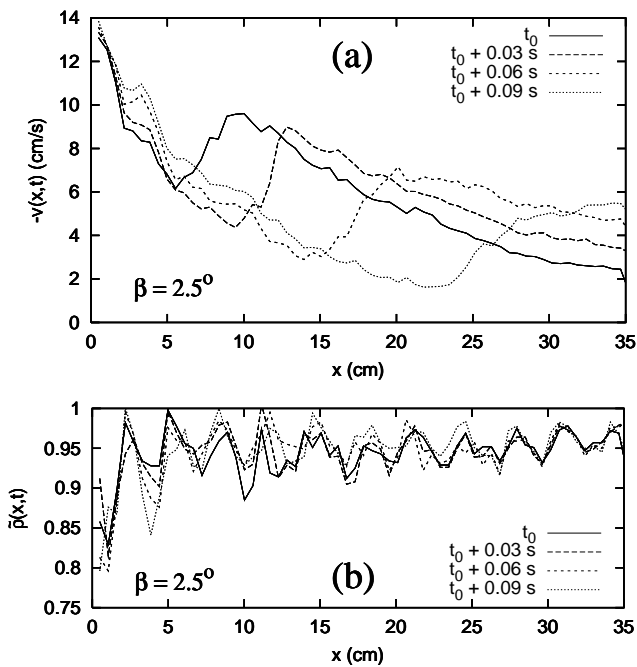}
}}
\caption{
The spatial dependence of (a) the velocity $-v(x,t)$ and (b) the
corresponding relative densities $\tilde{\rho}(x,t)$ for fixed $t$.
In these data, one is observing a decaying propagating shock.
}
\label{fig:12}
\end{figure}

Once again it it useful to look at the magnitudes of 
$v(x,t)$ and $\tilde{\rho}(x,t)$, but now for fixed $t$.
In Fig.~\ref{fig:12} we show $v(x,t)$ and $\tilde{\rho}(x,t)$ 
for four equally spaced consecutive times. 
In Fig.~\ref{fig:12}(a) one can see a shock propagating upstream 
with a speed $U \sim 250$~cm/s.
As already noted, it dies off as it moves upstream, but now
one can clearly see that the shock region also gets broader.
Fig.~\ref{fig:12}(b) indicates why the density $\tilde{\rho}(x,t)$ 
cannot be used to detect shock waves for large $\beta$. 
Except for the lowest 10~cm of the funnel, the density fluctuations 
are entirely dominated by the packing sites (which in this case have
a periodicity of 3.2~cm).
%%FIG_13
%% FIG 13:
\begin{figure}  
\epsfxsize=8.5cm
\centerline{\hbox{ 
\epsffile{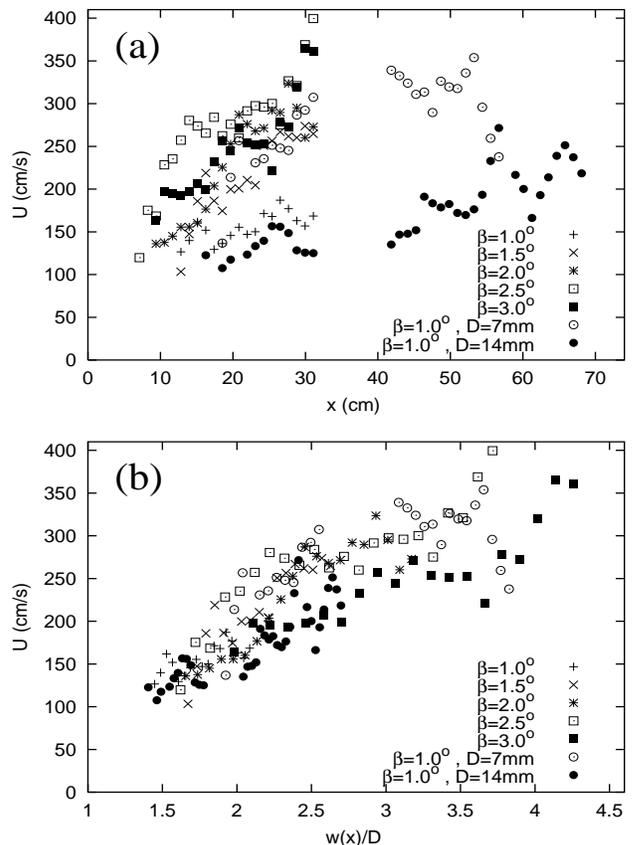}
}}
\caption{
(a) Estimated shock speed $U(x)$ for $1.0^{\circ} \leq \beta \leq 3.0^{\circ}$
and $D=7$, 10, and 14~mm.
(b) The same data replotted against the rescaled variable $w(x)/D$.
}
\label{fig:13}
\end{figure}

As already discussed in Sec.~\ref{profile}, it was observed in HD that 
for $0.1^{\circ} < \beta < 1.0^{\circ}$ the local average shock speed 
seemed to depend linearly on the local funnel width.
We can now test this hypothesis for larger $\beta$.
Unfortunately, with our current data constraints, we have only 
10-20 shocks so the statistics are not as good as could be desired.
Fig.~\ref{fig:13} shows $U(x)$ based on a simple ridge detection 
method applied to $a(x,t)$. 
Fig.~\ref{fig:13}(a) shows $U(x)$ for different values of $\beta$ 
and $D$ while (b) shows the same data versus $w(x)/D$. 
The data collapse in Fig.~\ref{fig:13}(b) is not complete,
although it is consistent with the same analysis in HD. 

\section{Shock creation} \label{shockstart}

Having established some understanding of how shock waves 
propagate, we would now like to examine under what
circumstances they are created. 
Since for $\beta > 0.1^{\circ}$ shock waves (once created) 
propagate all the way to the reservoir, the mechanisms of shock
creation will strongly effect the flow properties.
In HD it was found that for intermittent flows new shock waves were 
almost exclusively created at the packing sites $\chi_i$. 
Fig.~\ref{fig:04} illustrates how a moderately dense flow is
forced to reorganize at the packing sites, thus possibly increasing 
the likelihood of congestion there. 
%%FIG_14
%% FIG 14:
\begin{figure}  
\epsfxsize=8.5cm
\centerline{\hbox{ 
\epsffile{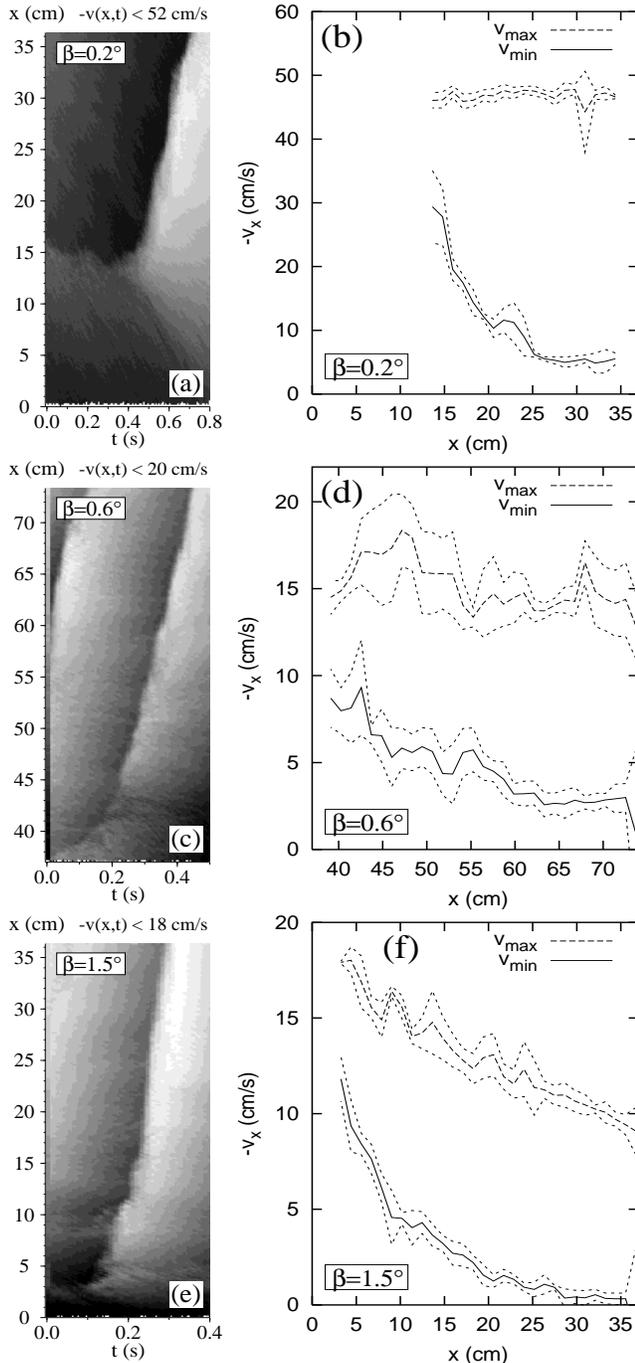}
}}
\caption{
Examples of starting shocks and their respective $v_{min}(x)$ 
and $v_{max}(x)$ for different funnel angles:
(a,b) $\beta = 0.2^{\circ}$, (c,d) $\beta = 0.6^{\circ}$, 
(e,f) $\beta = 1.5^{\circ}$. 
}
\label{fig:14}
\end{figure}

In Sec.~\ref{profile} it was discussed how $v_{min}$ varies in
the different flow types.
Basically $v_{min}$ is a measure of how effectively the shock has 
absorbed momentum and energy from the flow upstream of the shock.
Generally, we have found that $v_{min}$ for newly created shocks
is often significantly higher than the typical values of
$v_{min}(x)$ shown in Fig.~\ref{fig:10}(a) where
the shocks at A1 and A2 are known to be newly created.
This means that the new shocks do not dissipate energy as efficiently
as the older shocks around them.
Three examples of newly created shocks are shown in 
Figs.~\ref{fig:14}(a,c,e), and the averages of $v_{min}$ and 
$v_{max}$ of the balls encountering these shocks are shown 
in (b,d,f), respectively. 
One can observe how $v_{min}$ starts relatively high, but then over a distance
of 5-15~cm, it drops to the typical values found in Sec.~\ref{profile}. 
%%FIG_15
\begin{figure}  
\epsfxsize=8.5cm
\centerline{\hbox{ 
\epsffile{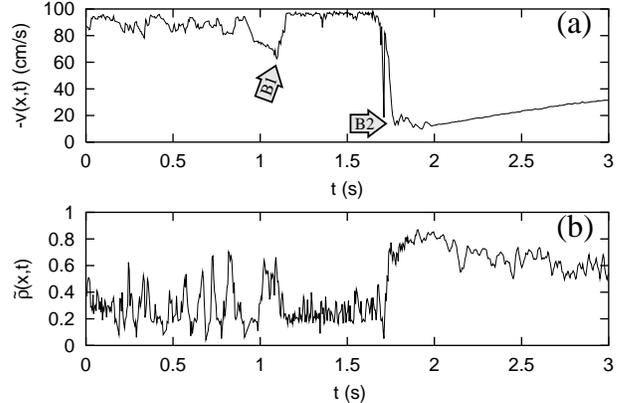}
}}
\caption{
Time series of (a) $-v(x,t)$ and (b) $\tilde{\rho}(x,t)$ 
at $x=30$~cm showing the creation of a shock in pipe flow 
($\beta = 0.075^{\circ}$). A slow, dense group of balls passes 
downstream (B1) and forms a shock at $x=26$~cm (not shown)
which starts propagating upstream (B2).
}
\label{fig:15}
\end{figure}

For each flow type we can make more specific 
observations about how shocks are typically created.
In pipe flow new shocks typically start as a group of 
balls moving downstream which have a somewhat smaller velocity 
and higher density than those around them.
This early stage of a shock at some point becomes stronger, 
stops, and then starts moving upstream as a stable shock
(shocks moving downstream are generally not very stable). 
An example of this process is shown in Fig.~\ref{fig:15} where
a disturbance passes the observation point at B1, turns around downstream,
and then starts propagating upstream, passing the observation 
point again at B2. 
Mass conservation across the shock discontinuity requires that 
\begin{equation} \label{eq:massconservation}
U = ( \rho_u v_u - \rho_d v_d )/( \rho_u- \rho_d )
\end{equation}
where the subscripts $u$ and $d$ signify 
upstream and downstream of the shock, respectively.  
In this example, $\rho_u \sim 0.3 w$, 
$\rho_d \sim 0.6 w \rightarrow 0.8 w$, and $v_u \sim -95$~cm/s. 
Note that as $v_d$ increases from $-70$~cm/s to $-20$~cm/s,
and $\rho_d$ increases slightly, $U$ changes sign. 

In intermittent flow the shock may start in the manner described above
for pipe flow, but more commonly it is created at a packing site
without the early downstream phase discussed above.
An example of such a shock is shown in  Fig.~\ref{fig:14}(a). 
\end{multicols}
\widetext
\begin{figure}  
\epsfxsize=17.5cm
\centerline{\hbox{ 
\epsffile{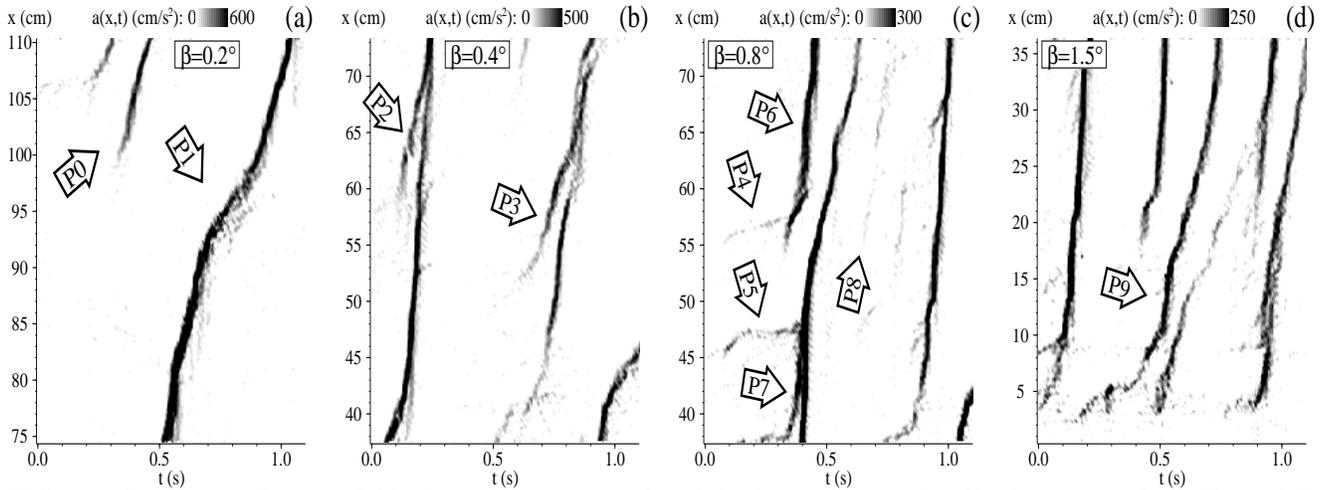}
}}
\caption{
Positive acceleration field $a(x,t)$ showing examples of 
shock interactions.
(a) Shock start repulsion (P0 and P1) ($\beta=0.2^{\circ}$). 
(b) Two temporary pre-shocks (P2 and P3) ($\beta=0.4^{\circ}$). 
(c) Stationary shocks joining at packing sites (P4 and P5), 
and a near shock repulsion (P6),
pre-shock (P7), and micro-shock (P8) ($\beta=0.8^{\circ}$).
(d) Near shock repulsion (P9) ($\beta=1.5^{\circ}$).
}
\label{fig:16}
\end{figure}
\begin{multicols}{2}
\narrowtext
In Fig.~\ref{fig:14}(b) we see how the growing difference
between $v_{min}$ and $v_{max}$ shows how shocks improve their ability 
to absorb momentum and energy during the first 10~cm of propagation.
Other examples of shock creation in intermittent flows have already
been shown in Figs.~\ref{fig:03}(a) and~\ref{fig:10}(a). 

In dense flow shocks are often created by the joining of a 
few relatively stationary weak ``pre-shocks'' 
(which will be discussed in the next section)
or a region with seemingly ``unstructured'' disturbances in the 
flow at a packing site (see, e.g., Figs.~\ref{fig:09} 
and \ref{fig:11}). 
The nature of these shock creation processes is not known.
Fig.~\ref{fig:14}(c) shows the creation of a shock wave 
38~cm upstream of the outlet. 
The $|v_{min}|$ shown in Fig.~\ref{fig:14}(d)
initially has a value somewhat higher than what is typical for that 
funnel position (see Fig.~\ref{fig:08}(a)).
In Fig.~\ref{fig:14}(e) the creation of a shock near the outlet
is shown for large $\beta$.  
Fig.~\ref{fig:14}(f) shows how its
$|v_{min}|$ also quickly drops after the creation.
Near the outlet (and one packing site up) all shocks have just been created. 
Therefore, the $v_{min}$ curve in Fig.~\ref{fig:14}(f) partially
corresponds to the first part of the curves in Fig.~\ref{fig:08}.
This also explains why the values of $v_{min}$ in 
Fig.~\ref{fig:10}(d) at $x=4.5$~cm are relatively high.

\section{Shock interactions} \label{interact}

Shock wave interactions have been discussed in HD.
Since these were studied using directly measured density 
fluctuations, they were hard to resolve and the nature of the 
interactions was somewhat speculative. 
The superior spatial and temporal resolution of the ball tracking 
method now gives us a clearer picture of the interactions 
and even reveals new interactions not observed earlier.

An example of a weak {\it shock start repulsion} is shown
in Fig.~\ref{fig:16}(a).
First, a new shock is created at point (P0). 
When the balls leaving this shock reach the next shock,
they cause it to temporarily slow down (P1), in effect repelling it.
(A similar patch of slower balls was observed in 
Fig.~\ref{fig:10}(a) at A3 and A4 following the creation 
of new shocks at A1 and A2.)

In  Fig.~\ref{fig:16}(b) examples of {\it pre-shocks} (P2 and P3) 
separated from the main shock by less than $0.1$~s are shown.
At P3 the pre-shock seems to gain amplitude and swallows the 
original main shock. 
Fig.~\ref{fig:16}(c) shows weak stationary shocks 
(P4 and P5) waiting to be swallowed by arriving shocks at 
or just upstream of packing sites. 
Examples are shown of a {\it near shock repulsion} at P6 and 
of a weak pre-shock at P7. 
Some {\em micro-shocks} also seem to propagate 
upwards with speeds similar to the surrounding shocks at P8. 
It is not clear what these represent, 
but since the balls are densely packed at $\beta=0.8^{\circ}$,
they may be ``chain collisions'' propagating in single rows of balls. 
The data set in Fig.~\ref{fig:16}(c) partially overlap the
data presented in Fig.~\ref{fig:09}.
In Fig.~\ref{fig:16}(d) a {\it near shock repulsion} is
seen at P9 and several weak {\it waiting shocks} are observed.
These waiting shocks seem related to the shock creation process
in dense flows discussed in Sec.~\ref{shockstart}.
There is no indication that shock interactions
disappear at larger $\beta$ (compare with Fig.~\ref{fig:11}).   

\section{Shock temperature} \label{grantemp}

When a ball encounters a shock, it quickly loses most of its velocity 
(and therefore most of its momentum and energy) in collisions 
with other balls and the funnel walls. 
In the dense region of the shock where the mean distance 
between balls is small, this results in relatively high collision 
rates when they encounter a fast incoming ball.
The high density also forces balls to come into contact with the
funnel walls so that the {\em total} momentum/energy of a
dense group of balls is also reduced by friction. 
This phenomenon is reminiscent of the clustering\cite{goldhirsch} 
and inelastic collapse\cite{mcnamara} observed in simulations 
of granular materials.

This process can also be described as follows.
Upstream of the shock there is a steady flow of relatively high energy balls.
When a group of balls encounters the shock, collisions ``randomize''
their translational energy, increasing the granular temperature. 
However, it is then reduced by each collision and by
the friction processes mentioned above. 
Some distance downstream of the shock the relative motion of 
neighboring balls appears to be negligible.
Thus, since we cannot measure collision rates in a shock,
the rate at which the granular temperature decays in a 
group of balls following the passage of a shock front
seems the best way of studying the energy loss processes in the shock. 
Consequently, we define $\tau_T$ as the characteristic decay time 
of $T(x,t)$ following the passage of a shock front 
in Lagrangian coordinates. 

%%FIG_17
%% FIG 17:
\begin{figure}  
\epsfxsize=8.5cm
\centerline{\hbox{ 
\epsffile{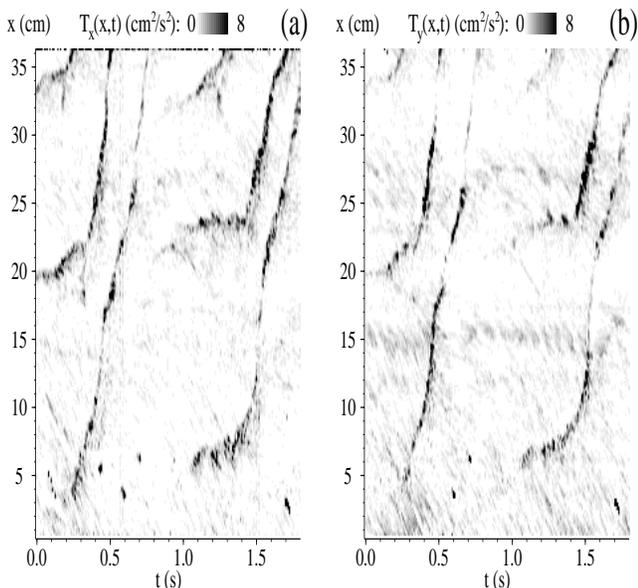}
}}
\caption{
Comparison of (a) $T_x(x,t)$ and (b) $T_y(x,t)$ showing creation
and propagation of shocks for $\beta = 0.6^{\circ}$.
}
\label{fig:17}
\end{figure}

First, we must consider the usefulness of $T(x,t)$ in the shock region.
It may be argued that since $v(x,t)$ is changing rapidly in the 
shock region, a calculation of $T_x(x,t)$ (and thus $T(x,t)$) 
is of limited value here. 
The coarse-grain resolution on which $T_x(x,t)$ is based is
$\lesssim 1$~ball diameter, while the shock region is typically
$\gtrsim 3$~ball diameters. 
Thus the resolution of $T_x(x,t)$ should suffice.  
In Fig.~\ref{fig:17} we compare $T_x(x,t)$ and  
$T_y(x,t)$ and find that in general they have the same magnitude,
except for some small packing site related oscillations.
(There is obviously less room for transverse movement of individual balls
near packing sites.)
Consequently, we will only consider $T(x,t)$ rather than
the individual components.
  
Transforming $T(x,t)$ to Lagrangian coordinates to study
the magnitude of $\tau_T$ would be difficult. 
Instead the following observation can be made. 
The characteristic decay {\em length} $\lambda_T$ of $T(x,t)$ downstream 
of the shock can be estimated from graphs of $T(x,t)$ for fixed $t$. 
Following the passage of a shock front by a group of balls, 
the shock will move upstream with a speed $U$ while the balls
move downstream with a speed  $\sim |v_{min}|$. 
Consequently, we find that $\tau_T \approx \lambda_T/(U+|v_{min}|)$.
%%FIG_18
%% FIG 18:
\begin{figure}  
\epsfxsize=8.5cm
\centerline{\hbox{ 
\epsffile{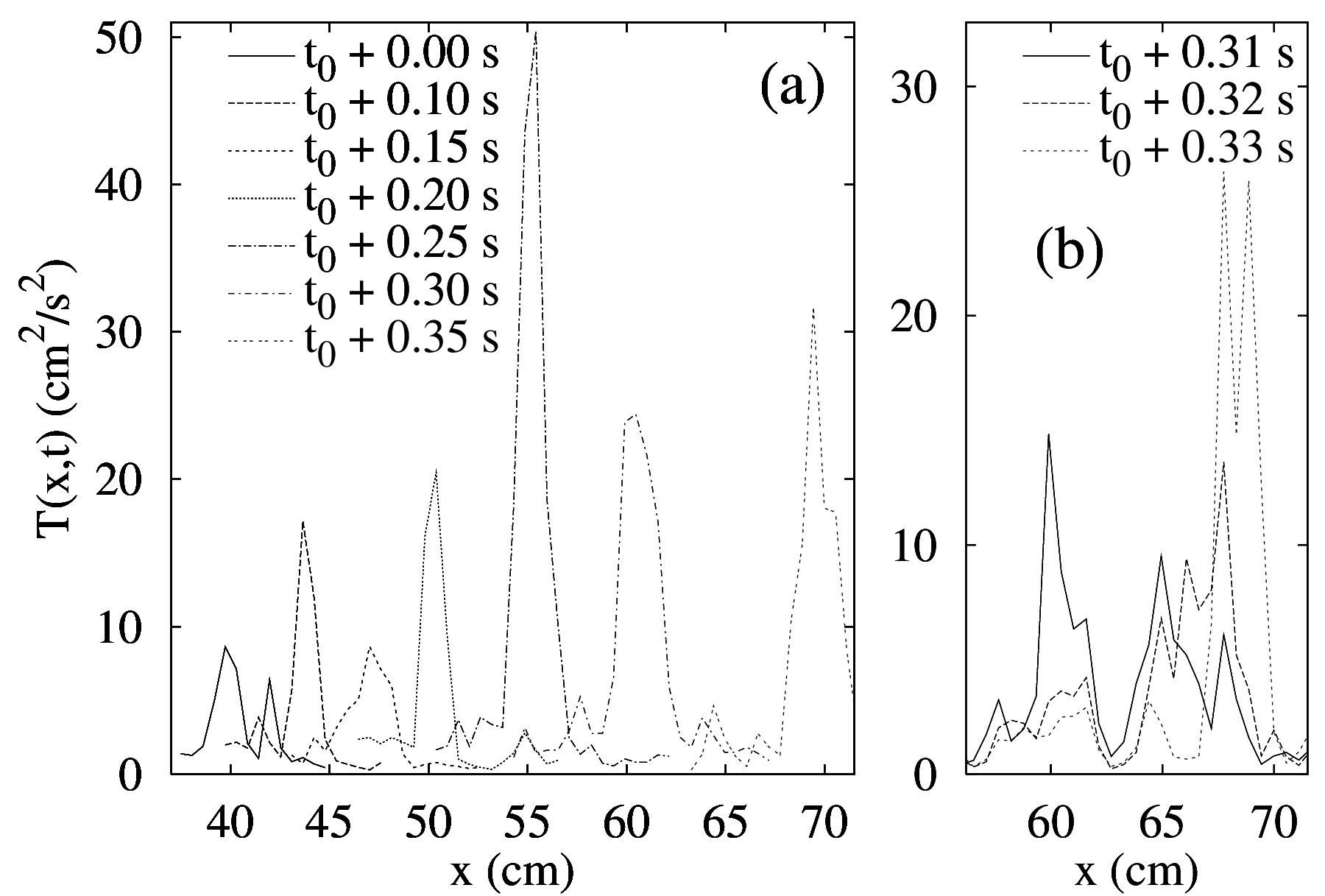}
}}
\caption{
Spatial dependence of $T(x,t)$ at the position of a propagating shock 
at subsequent times. In (a) the shock is created at $x=40$~cm 
($\chi_5 = 41$~cm) just before $t_0$. 
>From $t_0 + 0.1$~s to $t_0 + 0.35$~s the shock propagates from $x=44$~cm
to $x=70$~cm with increasing speed. 
In (b) the passage of the shock at the next packing site ($\chi_6=67$~cm) 
is shown with better time resolution.
}
\label{fig:18}
\end{figure}

The spatial dependence of $T(x,t)$ around a propagating shock is shown 
in Fig.~\ref{fig:18} for a series of fixed times.
In Fig.~\ref{fig:18}(a) a shock wave is created at $\chi_5=41$~cm
and propagates from $x=40$~cm to $x=70$~cm in $0.35$~s. 
The propagation past $\chi_6=67$~cm is shown with higher temporal
resolution in Fig.~\ref{fig:18}(b). 
Due to the dense packing at and just below a packing site 
($62$~cm $\leq x \leq 67$~cm in Fig.~\ref{fig:18}(b)), 
there is little room for relative motion between the balls. 
When a shock passes this region, $T(x,t)$ remains small
but has peaks both upstream and downstream. 
The tendency of $T(x,t)$ to remain small in areas just below 
packing sites is generally observed 
(see Figs.~\ref{fig:09}(d) and \ref{fig:17}).
For large funnel angles ($\beta > 1^\circ$), the region of increased 
temperature associated with a shock may extend over the 
intervals between three or more packing sites.    
From Fig.~\ref{fig:18}(a) and(b) we estimate that
$\lambda_T \sim 2-5$~cm, hence $\tau_T \lesssim 0.04$~s. 

From plots of $T(x,t)$ it is easy to estimate the decay rate of
$T(x,t)$ in Eulerian coordinates $\tau_T^{\ast} \approx \lambda_T/U$.  
When $U >> |v_{min}|$, we find that $\tau_T^{\ast} \approx \tau_T$.
Figs.~\ref{fig:09}(d) and \ref{fig:17} yield the estimate
$\tau_T^{\ast} \lesssim 0.05$~s. 
Generally, it seems that $\tau_T^{\ast}$ is always significantly
shorter than the characteristic time separation between neighboring
shocks and thus fluctuations in the flow velocity created by one shock 
generally do not reach the next shock.

A more extended region of increased $T(x,t)$ is often observed
where new shocks are created 
(see Figs.~\ref{fig:09}(d) and \ref{fig:17}). 
This may be due to the less efficient packing of balls in these shocks 
(which may be related to the inefficient energy dissipation and large 
values of $|v_{min}|$ discussed in Sec.~\ref{shockstart}).

In the above discussion of magnitudes and decay rates of $T(x,t)$ it
should be recalled that $T(x,t)$ is based on measurements of $v_x$ and
$v_y$ with time resolutions $\sim 0.02$~s (see Sec.~\ref{trackingmethod}). 
Granular temperature associated with very high collision rates 
therefore cannot be resolved. 
Due to these high collision rates, the unmeasured part of the 
granular temperature must decay even faster than the measured part,
and consequently better time resolution would merely increase the 
magnitude of $T(x,t)$ in the central shock region and not change 
the observed decay rates.

\section{Average properties}\label{xavg}
%%FIG_19
%% FIG 19:
\begin{figure}  
\epsfxsize=8.5cm
\centerline{\hbox{ 
\epsffile{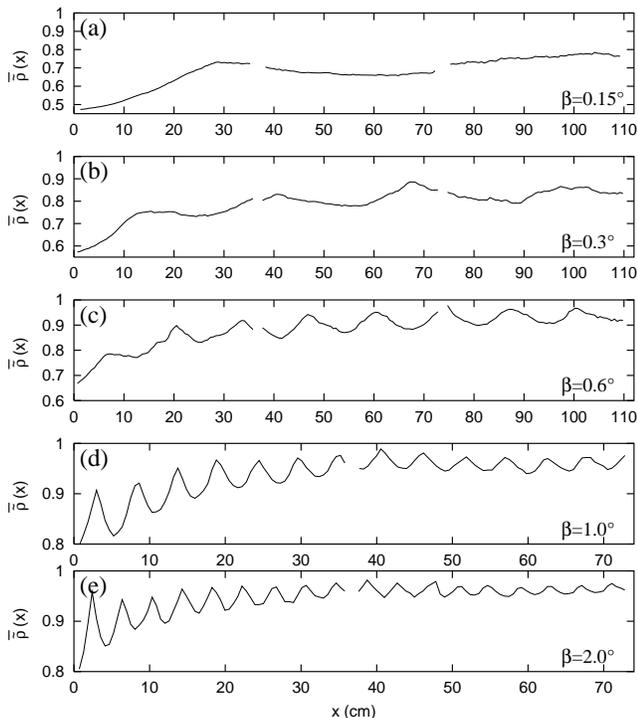}
}}
\caption{
Spatial dependence of the time averaged relative density
$\bar{\tilde{\rho}}(x)$ showing the packing site periodicity 
and decay near the outlet for different funnel angles:
(a) $\beta=0.15^{\circ}$, (b) $\beta=0.3^{\circ}$,
(c) $\beta=0.6^{\circ}$, (d) $\beta=1.0^{\circ}$,
(e) $\beta=2.0^{\circ}$. 
(All curves are composed of two or three data sets, hence, there
are some gaps in the data.)
}
\label{fig:19}
\end{figure}

Moving beyond the flow behavior associated with individual shocks,
we will now study the spatial dependence of the time averaged values
of $\tilde{\rho}(x,t)$ and $v(x,t)$ which we denote as
$\bar{\tilde{\rho}}(x)$ and $\bar{v}(x)$ respectively. 
It is important to remember that the measurements only
cover a time span corresponding to 3-15 passing shock waves. 
It was established earlier that individual shock waves strongly
influence $\tilde{\rho}(x,t)$ and $v(x,t)$. 
Thus, measurements with only a few shocks (usually intermittent flows) 
may exhibit large fluctuations from the ``true'' mean values.
By combining the averages of these two or three measurements from
contiguous sections of the funnel, we obtain the statistics 
in the lowest 110~cm of the funnel in the measurements presented below.

In Fig.~\ref{fig:19} we show $\bar{\tilde{\rho}}(x)$ for various 
intermittent and dense flows. 
In the intermittent flows in Figs.~\ref{fig:19}(a) and (b), 
we see that $\bar{\tilde{\rho}}(x)$ has a slow growing trend 
throughout the observed region, with weak packing site related oscillations.
For the dense flows in Figs.~\ref{fig:19}(c-e), 
$\bar{\tilde{\rho}}(x)$ is nearly unity except near the outlet,
and packing site effects are more pronounced. 
The general decrease of $\bar{\tilde{\rho}}(x)$ near the outlet 
is most likely caused by the passing of fewer shock waves there.
%%FIG_20
%% FIG 20:
\begin{figure}  
\epsfxsize=8.5cm
\centerline{\hbox{ 
\epsffile{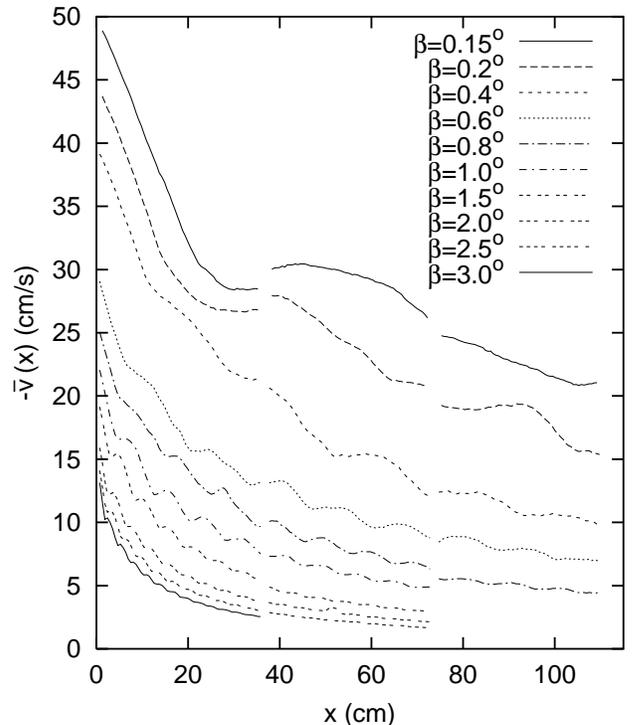}
}}
\caption{
Spatial dependence of the time averaged ball velocities $-\bar{v}(x)$ 
showing weak packing site periodicity and a decreasing overall dependence 
on $x$.
(All curves are composed of two to three data sets, hence, there
are gaps in the data.)
}
\label{fig:20}
\end{figure}

In Fig.~\ref{fig:20} the average velocities are shown for different 
values of $\beta$.
For $\beta < 0.4^{\circ}$ the statistics are not quite good enough 
to ensure that the curves fit together. 
In all curves a general decreasing trend is observed,
on top of which there are weak packing site related effects.
Average ball speeds are lowest at the packing sites. 
Since the average flow rate is constant throughout the funnel, 
then $|\bar{v}(x)| \sim 1 / w(x) \bar{\tilde{\rho}}(x)$. 
For large $\beta$, $\bar{\tilde{\rho}}(x)$ is essentially constant 
and thus $ |\bar{v}(x)| \sim 1/w(x)$.
%% FIG 21:
\begin{figure}  
\epsfxsize=8.5cm
\centerline{\hbox{ 
\epsffile{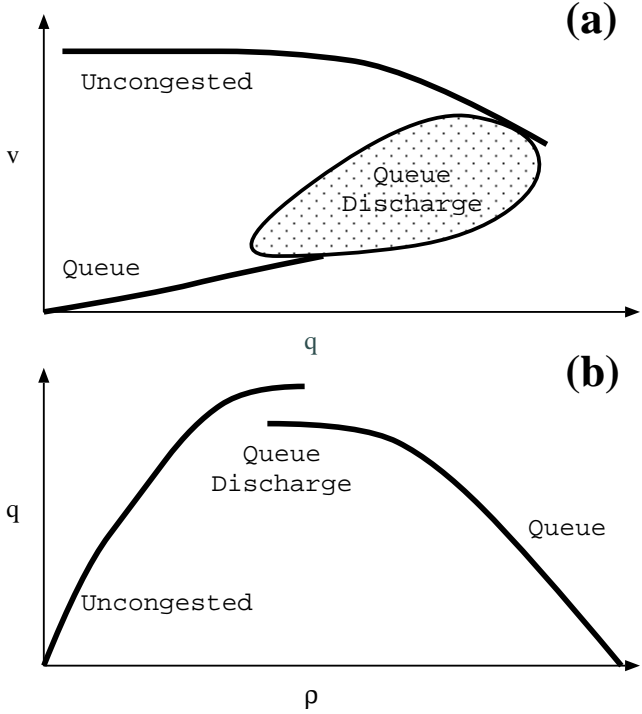}
}}
\caption{
Schematic drawings of the (a) flow/speed diagram $v(q)$ and the
(b) fundamental diagram $q(\rho)$.
}
\label{fig:21}
\end{figure}

\section{Comparison with traffic flow} \label{traffic}

Three flow types have been identified in traffic flow, 
namely, {\em uncongested flow} (steady flow of vehicles 
with speeds $\sim 100$~km/h and low to moderate densities), 
{\em queue flow} (slow flow from 0-20~km/h and near maximum density), 
and {\em queue discharge} (accelerating vehicles leaving
a queue flow situation)\cite{wolf}.
Traffic data are usually presented as $v(q)$ 
(the speed/flow relation) or as $q(\rho)$ (the fundamental diagram), where 
$v$, $q$, and $\rho$ are the velocity, flow rate, and density, respectively.
As shown schematically in Fig~\ref{fig:21}, both $v(q)$ (a) and 
$q(\rho)$ (b) have regions associated with the three 
flow types mentioned above.
The fundamental diagram often resembles an inverted 
parabola\cite{wolf,lighthill,whitham}, which may contain a 
discontinuity near its maximum as shown in Fig.~\ref{fig:21}(b).

\begin{figure}  
\epsfxsize=17.5cm
\centerline{\hbox{ 
\epsffile{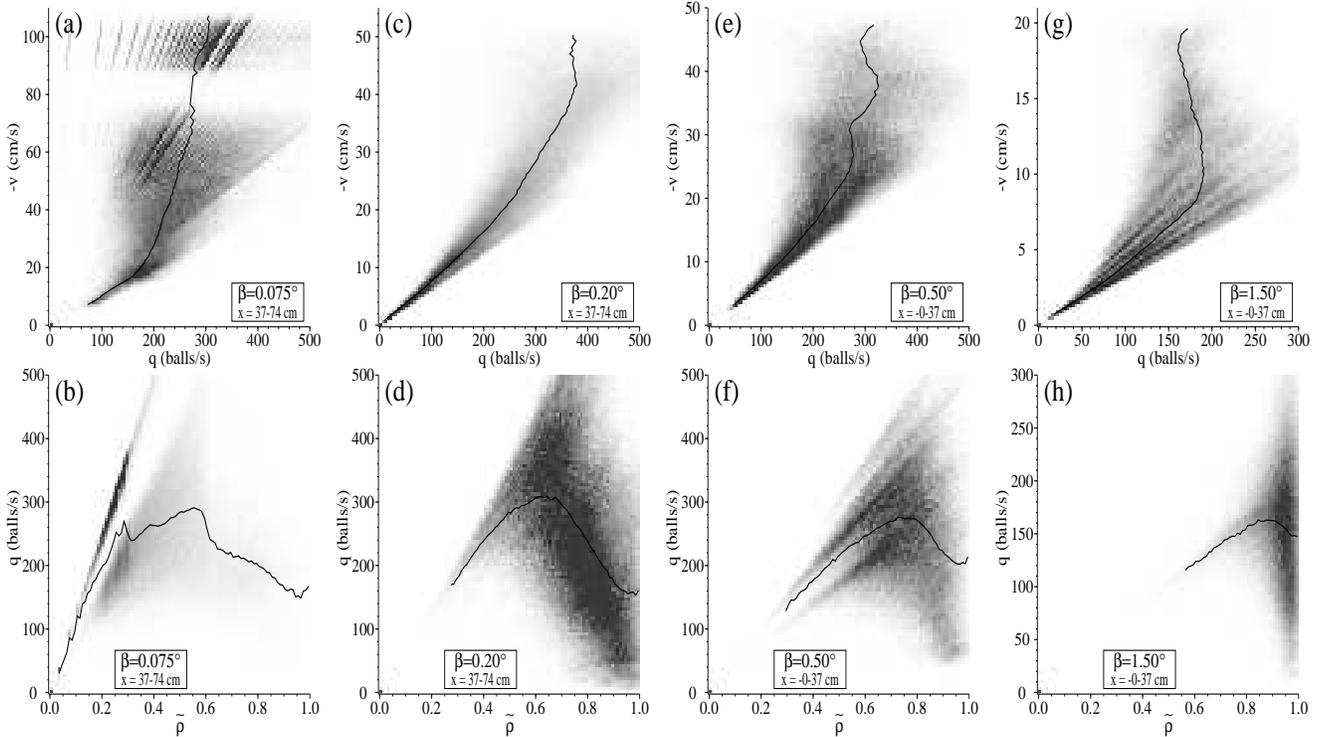}
}}
\caption{
The $(q,v)$ and $(\rho,q)$ distributions for four different funnel angles:
(a,b) $\beta=0.075^\circ$, (c,d) $\beta=0.20^\circ$,
(e,f) $\beta=0.50^\circ$, (g,h) $\beta=1.50^\circ$.
The solid lines are the averages, indicating the flow/speed diagram $v(q)$
or the fundamental diagram $q(\tilde{\rho})$ as appropriate.
}
\label{fig:22}
\end{figure}

In Fig.~\ref{fig:22} we show grey scale histograms of $(q,v)$ and 
$(\tilde{\rho},q)$ for four different values of $\beta$. 
The average values, corresponding to $v(q)$ and $q(\tilde{\rho})$,
are shown as solid lines.
(Note that we use the relative density $\tilde{\rho}$ instead of $\rho$.)
As noted in Sec.~\ref{profile}, at any given time the balls in our system 
are either nearly motionless in a shock wave (corresponding to queue flow) 
or slowly speeding up (corresponding to queue discharge). 
No regimes corresponding to steady high speed flow 
(i.e., uncongested flow) were found for $\beta > 0.05^\circ$. 
\end{multicols}
\widetext
\begin{multicols}{2}
\narrowtext

For pipe flow ($\beta=0.075^{\circ}$)
Fig.~\ref{fig:22}(a) shows a small queue region and then  
a large region similar to queue discharge behavior. 
In the fundamental diagram in Fig.~\ref{fig:22}(b) 
low densities dominate. 
The $q(\tilde{\rho})$ curve in (b) (solid line) has a parabolic shape 
which peaks at $\tilde{\rho} \sim 0.5$. 
The data set from which the distributions were derived contained only one 
shock wave, hence, certain regions contain no data since some flow states 
(e.g., $v \sim 80$~cm/s) simply did not occur during the measurement.

For intermittent flow ($\beta = 0.2^{\circ}$) Fig.~\ref{fig:22}(c) 
displays a longer queue flow region than Fig.~\ref{fig:22}(a).
It then changes smoothly into a relatively broader queue discharge region.
The $(\tilde{\rho},q)$ distribution in Fig.~\ref{fig:22}(d) 
is dominated by moderately large densities ($\tilde{\rho} > 0.5$)
and $q(\tilde{\rho})$ peaks at a slightly higher value 
($\tilde{\rho} \sim 0.7$) than pipe flow.

For the denser flows shown in Figs.~\ref{fig:22}(e) and (g), 
it becomes increasingly difficult to distinguish between
queue flow and queue discharge.
The queue discharge regions are found at lower values
of $v$ and $q$, which is consistent with the general trend of falling average 
velocities (see Sec.~\ref{xavg}) and average flow rates 
($q \sim \beta^{-0.4}$ for $\beta > 0.5^\circ$\cite{VD96}).
The corresponding fundamental diagrams in 
Figs.~\ref{fig:22}(f) and (h) again have parabolic shapes
which peak at increasingly higher densities ($\tilde{\rho}> 0.8$).
The shapes of the distributions close to
$\tilde{\rho} = 1 $ are not very well-defined.
We know from Secs.~\ref{highbeta} and~\ref{xavg} that virtually 
all density variations are related to packing site effects at large $\beta$. 
Consequently, it should not be expected that the distribution for 
$\tilde{\rho} >0.9$ can be resolved.
For $\tilde{\rho}<0.9$, the behavior in Fig.~\ref{fig:22}(h) is most 
likely caused by the manner in which the dense flow leaves the outlet.

In real traffic, shock waves are generally slower than 
vehicle speeds (see~\cite{whitham}) and the densities are 
low ($\tilde{\rho} < 0.3$) during uncongested flow. 
These properties are similar to the flow conditions in true pipe 
flow ($\beta = 0^\circ$), whereas our few measurements 
(of duration $\sim 8$~s) do not have the time and space resolution
necessary to catch the full range of the shock dynamics.  
Thus we cannot display ball flow behavior corresponding to uncongested flow.  

Flows for $\beta > 0.2^{\circ}$ are strongly affected 
by packing sites and thus by the changing number of {\it lanes} 
as illustrated in Fig.~\ref{fig:04}. 
Several consecutive reductions in the number of lanes in traffic 
are seldom found, but a comparison would be interesting. 
A more general comparison between the manner in which granular shock waves 
and traffic jams propagate in space and time would also be interesting, 
but unfortunately there are practically no available $v(x,t)$ measurements 
in traffic with the appropriate temporal and spatial resolution. 

\section{Summary} \label{summ}

We have presented the results of particle tracking measurements 
on a two-dimensional flow of monodisperse balls.
The circumstances of the creation and propagation of shock waves
have been studied.
In particular, we have observed how individual balls behave both in 
the shock region and between shocks, and established that the 
interaction of balls between shocks can be disregarded by a 
simple rescaling of gravity.
The simultaneous tracking of thousands of balls has been used
to study the time and space variations of quantities such as
flow velocity, acceleration and granular temperature.
This has been used to study the shock wave behavior in very dense 
flows, the processes surrounding the creation of new shocks, 
and the role played by granular temperature.                    

\acknowledgments
It is a pleasure to thank C. Veje for his assistance and insights and
V. Putkaradze and M. van Hecke for many informative discussions. 
P.D. would like to thank Statens Naturvidenskabelige Forskningsr{\aa}d 
(Danish Research Council) for support.

\appendix

\section{Ball position determination} \label{app:ballfind}

Each stored video frame is recalled from the harddisk as an array of 
unsigned 8-bit integers.
This array is equivalent to a grey scale image where black 
corresponds to 0 and white to 255.
Each frame is subtracted from the image of the empty funnel. 
Thus, the balls stand out as a white patch on a black background. 
We will call these patches {\it ball images}. 
Based on the local distributions of grey scale values, a threshold 
is determined.
Everything above the threshold is defined as part of a ball.
Areas of adjoining pixels with values over the threshold belong
to the same ball. An estimate of the ball center position 
$(x_c,y_c)_{est}$ (based on a weighted average of the pixel values) 
and height (intensity in grey scale) is made.
The height and estimated positions $(x_c,y_c)_{est}$ will be used 
as starting values in a fit of a ball's image to a rotationally symmetric 
``bell-shaped'' function representing it.

Next, we need to determine which pixels belong to the ball image
and should be used in the fit.
An area of  pixels surrounding the estimated ball center 
$(x_c,y_c)_{est}$ is selected.
Let $d$ be the distance from $(x_c,y_c)_{est}$
for each pixel and let $r$ be the ball radius. 
Pixels with $d \leq 0.9 r$ are all selected. 
Pixels with $0.9 r < d < 1.45 r$ are selected if the 
next pixels away from  $(x_c,y_c)_{est}$ are not higher than
the value of the pixel in question (if they are, it indicates
that the value of the pixel is influenced by a wall or another ball).
All other pixels are disregarded. 
In this way, most of the pixels belonging to the ball image
and some of the surrounding black ones will be selected,
while pixels belonging to other balls will be left out.

Typically, between 20 and 35 pixels are selected for the position 
fit of each ball. 
These pixel values are fitted to a rotationally symmetric 
bell-shaped function with 4 fitting parameters: 
peak height, $(x,y)$ components of the center position, and the ball radius. 
(The convergence of the fits is more stable when the radius is 
allowed to vary.)
When a radial distribution of pixel values is found, the
resulting curve matches the function $e^{-{|d|}^3}$ 
remarkably well, where $d$ is the pixel distance from $(x_c,y_c)$. 
In most fits, a bell-shaped function with a radial 
dependence $\sim e^{-{|d|}^3} $ also gives the best results.
In a small number of fits ($<1\%$), however, a
stable fit is not achieved with this function. 
In these cases a second fit is made with an ordinary Gaussian
function $e^{-{|d|}^2}$ . 
The Gaussian fits give slightly inferior results on average 
but are much more stable in the ``difficult'' cases.
All fits were made using the {\it amoeba method}~\cite{numrec}.
At this point a table of ball center positions has been determined
for each frame in the sequence.
 
The error of the ball center positions ($\lesssim 0.15$~mm)
was determined in two ways.
In the first, a number ($\sim 20$) of balls were glued to a rigid,
transparent piece of plastic. This piece was translated and rotated
in the funnel and a number of frames were shot. In each frame the
relative positions of the balls should be constant and the noise
level could subsequently be determined.
In the second, individual balls were allowed to roll freely
in the funnel (without touching the funnel walls or other balls). 
>From the smoothness of enlargements of the resulting trajectories,
the noise level of the position detection could be estimated. 

\section{Interference with pixel array of camera} \label{app:pixelif}

\begin{figure}  
\epsfxsize=8.5cm
\centerline{\hbox{ 
\epsffile{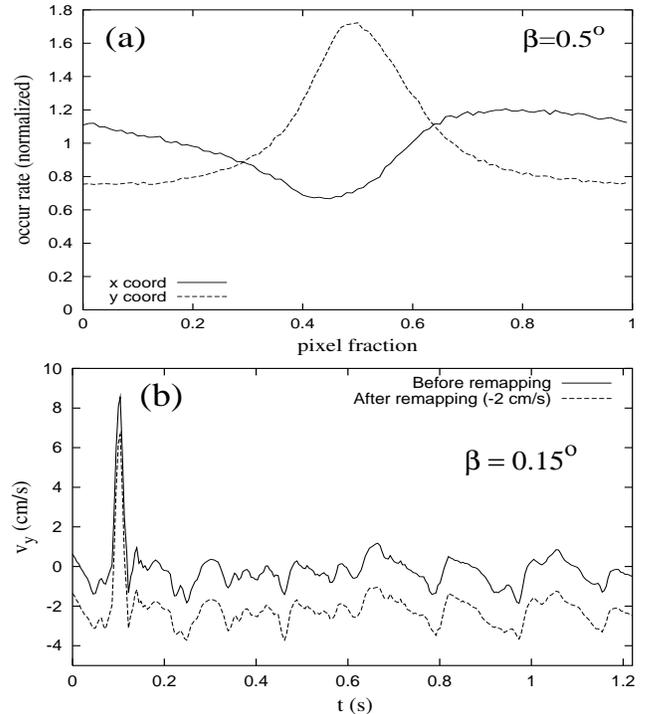}
}}
\caption{
(a) Distribution of fractional part of ball coordinates measured in
pixel coordinates ($\beta = 0.5^{\circ}$, $D = 10$mm, 
$\theta = 4.1^{\circ}$). 
(b) Comparison of $v_y$ for the same ball before and after re-mapping of
ball coordinates. The ``after re-mapping'' curve is displaced $-2$~cm/s. 
($\beta = 0.15^{\circ}$, $D = 10$mm, $\theta = 4.1^{\circ}$). 
(The ball encounters a shock at $t=0.1$~s.)
}
\label{fig:23}
\end{figure}

In original versions of ball center position histograms
of the type shown in Fig.~\ref{fig:04},
there were clear signs of horizontal stripe patterns where
the distance between the stripes corresponded to the distance
between neighboring lines of pixels in the camera. 
This indicated that there was some interference between
the pixel periodicity of the camera and the determined ball
center positions since there is no reason why some positions
relative to the pixel array should occur significantly more often
than others.  Subsequent histograms of the fractions of
the ball center positions in pixel coordinates were made.
An example of such a histogram is shown in Fig.~\ref{fig:23}(a). 
These histograms confirmed the existence of interference along
both the $x$ axis and the $y$ axis.
For simplicity, we assume that this interference was the same 
in all areas of the camera and that this systematic remapping
is ``smooth''. With these assumptions it is relatively easy
to map the positions back to their ``true'' values and thus
eliminate the effect of this interference. 
The resulting repositionings of the ball centers are less
than half the statistical uncertainty of the positions and
thus it is of limited importance. 
In Fig.~\ref{fig:23}(b) a comparison is made between the 
same ball trajectory before and after the remapping. 
As can be seen, the effect is relatively small.
The remapping has nevertheless been done for all data sets 
to avoid any ``cumulative'' effects of this systematic error. 
In Figs.~\ref{fig:04}(a,b) there is still a faint set 
of horizontal lines (with a vertical periodicity of $\sim 0.6$~mm)
but this has been significantly reduced by the remapping.   
%%FIG_23
%% FIG 23:

\section{Assumption of transverse uniformity} \label{app:yprof}

Throughout this article the flow of balls has been treated as
essentially one-dimensional. 
Since the width of the funnel ($\lesssim 10$~ball diameters)  
is relatively small compared with the other important length scales 
in the system (e.g., funnel length and shock separation),
it is reasonable to assume that the flow is uniform across the funnel.
Between shocks, balls essentially move independently of each other 
(see Sec.~\ref{profile}) and uniformity may be harder to maintain.
In any case, this assumption should be checked.
It is unfortunately hard to devise any meaningful statistics that could 
confirm the instantaneous transverse equilibrium of a moving shock wave. 
By averaging over $x$ and $t$ on the other hand, we can 
achieve meaningful statistics and subsequently check
the $y/w(x)$ dependence of various flow properties.
(When averaging over $x$ is done we must renormalize $y$ with
$w(x)$ to make the averaging meaningful.)
%%FIG_24
%% FIG 24:
\begin{figure}  
\epsfxsize=8.5cm
\centerline{\hbox{ 
\epsffile{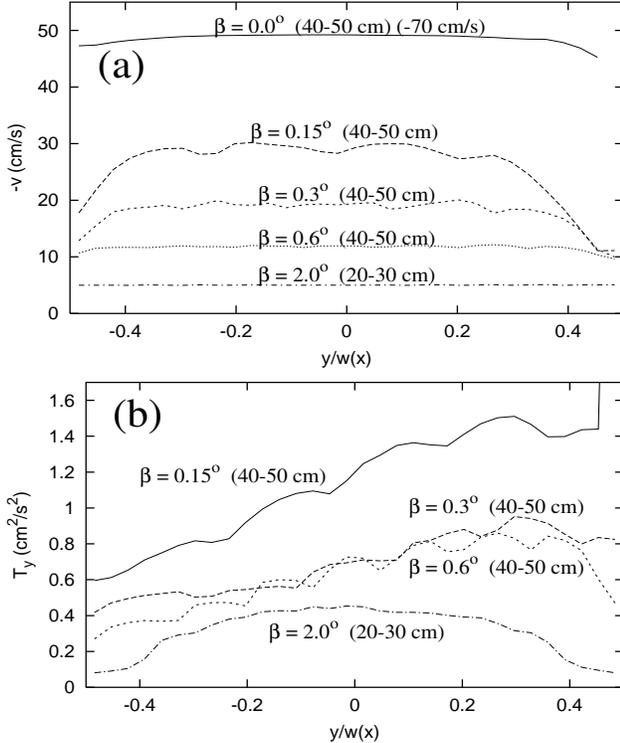}
}}
\caption{
The space and time averaged velocity profile versus $y/w(x)$
for different funnel angles ($D=10$~mm, $\theta=4.1^{\circ}$).  
}
\label{fig:24}
\end{figure}

In Fig.~\ref{fig:24} the $y/w$ dependence of $v$ is shown averaged 
over both $x$ (in 10~cm intervals) and $t$ for various flows. 
Both pipe flow and dense flow show flat velocity profiles while 
intermittent flow ($\beta=0.15^{\circ}$ and $\beta=0.3^{\circ}$) 
exhibits a drop near the funnel walls. 
This is {\em not} evidence of a shear flow situation. 
It simply means that between shocks (where high ball speeds
occur) the regions are partially ``statistically empty'', namely,
when balls leave a shock, they uniformly occupy the full width
of the funnel. The balls closest to the funnel walls will often
collide with the wall and move towards the center of the funnel.
In the middle part of the funnel they will collide with several 
other balls and in each collision the transverse energy 
will be reduced and the transverse momentum will be averaged out 
(since balls are coming from both walls).
As a result the density will be somewhat higher in the middle of the
funnel compared with regions close to the funnel walls.
Thus, in the overall statistics, the fastest balls mostly contribute
in the central part of the funnel. 
In pipe flow ($\beta=0^{\circ}$) this effect is not seen because
the density is too low to eliminate transverse energy in the center
of the funnel, since each ball basically follows its own ``zig-zag''
trajectory. (Note that the data for $\beta=0^{\circ}$ has been
shifted downwards for clarity.  The actual velocities are $\sim 120$~cm/s.)
In dense flows ($\beta=0.6^{\circ}$ and $\beta=2.0^{\circ}$) the
effect is not observed since the density is constantly high 
(see Figs.~\ref{fig:19}(c) and~(e)) and there is no space for balls 
close to the funnel walls to move away as they accelerate following a shock.

\end{multicols}

\widetext

\newpage

\setlength{\unitlength}{1cm}

%% FIG 22 (Double column):

\end{document}